\documentclass{aastex701}

\newcommand{\objname}{SMDET-1}
\newcommand{\planetarium}{Plan\'etarium de Montr\'eal, Espace pour la Vie, 4801 av. Pierre-de Coubertin, Montr\'eal, Qu\'ebec, Canada}
\newcommand{\irex}{Trottier Institute for Research on Exoplanets, Universit\'e de Montr\'eal, D\'epartement de Physique, C.P.~6128 Succ. Centre-ville, Montr\'eal, QC H3C~3J7, Canada}

\begin{document}

\title{\objname: a Nearby Y Dwarf Candidate}

\author[orcid=0000-0002-1125-7384]{Aaron M. Meisner}
\affiliation{NSF National Optical-Infrared Astronomy Research Laboratory, 950 North Cherry Avenue, Tucson, AZ 85719, USA}
\affiliation{Carl and Lily Pforzheimer Foundation Fellow, Radcliffe Institute for Advanced Study at Harvard University, 10 Garden Street, Cambridge, MA 02138, USA}
\affiliation{Center for Astrophysics $|$ Harvard \& Smithsonian, 60 Garden St., Cambridge, MA 02138, USA}
\email[show]{aaron.meisner@noirlab.edu}  

\author[0000-0001-7896-5791]{Dan Caselden}
\email{DanCaselden@gmail.com}
\affiliation{Department of Astrophysics, American Museum of Natural History, Central Park West at 79th Street, NY 10024, USA}

\author[0000-0001-7519-1700]{Federico Marocco}
\email{federico@ipac.caltech.edu}
\affiliation{IPAC, Mail Code 100-22, Caltech, 1200 E. California Blvd., Pasadena, CA 91125}

\author[0000-0003-4269-260X]{J. Davy Kirkpatrick}
\email{davy@ipac.caltech.edu}
\affiliation{IPAC, Mail Code 100-22, Caltech, 1200 E. California Blvd., Pasadena, CA 91125}

\author[0000-0002-2592-9612]{{Jonathan Gagn\'{e}}}
\email{jonathan.gagne.1@gmail.com}
\affiliation{\planetarium}
\affiliation{\irex}

\author[0000-0002-6294-5937]{Adam C. Schneider}
\email{aschneid10@gmail.com}
\affiliation{United States Naval Observatory, Flagstaff Station, 10391 West Naval Observatory Rd., Flagstaff, AZ 86005, USA}

\author[0000-0002-6721-1844]{Samuel A. Beiler}
\email{BEILERS@tcd.ie}
\affil{School of Physics, Trinity College Dublin, The University of Dublin, Dublin 2, Ireland}

\author[0000-0001-9309-0102]{Sergio B. Fajardo-Acosta}
\email{fajardo@ipac.caltech.edu}
\affiliation{IPAC, Mail Code 100-22, Caltech, 1200 E. California Blvd., Pasadena, CA 91125}

\author[0000-0001-6251-0573]{Jacqueline K. Faherty}
\email{}
\affiliation{Department of Astrophysics, American Museum of Natural History, Central Park West at 79th Street, NY 10024, USA}

\author[0000-0003-3050-8203]{Stanimir A. Metchev}
\affiliation{Western University, Department of Physics and Astronomy, London, Ontario, Canada}
\affiliation{Western University, Institute for Earth and Space Exploration, London, Ontario, Canada}
\email{smetchev@uwo.ca}

\author[]{Sam Barber}
\affiliation{Western University, Department of Physics and Astronomy, London, Ontario, Canada}
\email{sbarbe24@uwo.ca}

\author[0000-0002-2387-5489]{Marc J. Kuchner}
\email{}
\affil{Exoplanets and Stellar Astrophysics Laboratory, NASA Goddard Space Flight Center, 8800 Greenbelt Road, Greenbelt, MD 20771, USA}

\author[0000-0003-2235-761X]{Thomas P. Bickle}
\email{tombicklecrypto@gmail.com}
\affiliation{School of Physical Sciences, The Open University, Milton Keynes, MK7 6AA, UK}

\author[0000-0003-4408-0463]{Zafar Rustamkulov}
\email{zafar@ipac.caltech.edu}
\affiliation{IPAC, Mail Code 100-22, Caltech, 1200 E. California Blvd., Pasadena, CA 91125}

\collaboration{all}{The Backyard Worlds: Planet 9 Collaboration}

\begin{abstract}

We present the discovery of \objname, a red, fast-moving object ($\mu \approx 1\farcs3$/yr) identified in time-resolved unWISE coadds using a pixel-level deep learning methodology called SMDET. Despite being relatively bright at 4.5~$\mu$m compared to many other recent WISE-based brown dwarf discoveries ($m_{[4.5]} \approx 14.6$ mag Vega), \objname~had remained overlooked due to its location in a very crowded Galactic plane field ($b \approx 2.25^{\circ}$) and contamination from brighter background objects. \objname~is also serendipitously detected at 4.5~$\mu$m in late-2012 Spitzer Deep GLIMPSE survey imaging. \objname~is undetected in UKIDSS and Palomar/WIRC near-infrared imaging, with the strongest constraint on its temperature ($T_{\rm eff} < 391$~K) arising from its Deep GLIMPSE color limit of $m_{[3.6]} - m_{[4.5]} > 2.81$ mag, which also implies a very nearby photometric distance $< 7.4$ pc. The Spitzer color bound corresponds to a Y dwarf phototype. \objname~illustrates the importance of continued searches for nearby brown dwarfs within archival datasets like WISE and Spitzer, as well as the potential of pixel-level deep learning to discover astronomical moving objects that challenge traditional data analysis approaches.

\end{abstract}


\section{Introduction} \label{sec:intro}

Y dwarfs \citep[$T_{\rm eff} \lesssim 500$~K;][]{Cushing_2011,Kirkpatrick_2011,Kirkpatrick_2012}, the coldest known spectral class of brown dwarfs, are the key to understanding the bottom of the field substellar mass function \citep[e.g.,][]{Leggett_2019b,Kirkpatrick_2021a,Kirkpatrick_2024}. However, due to the incompleteness of the identified Y dwarf sample even within the very local $d < 20$~pc volume, it remains impossible to pin down the low-mass cutoff of star formation, if one exists \citep{Kirkpatrick_2024}. Completing the nearby Y dwarf census remains a major challenge. Options for doing so include continued mining of archival 3-5~$\mu$m data sets from the Wide-field Infrared Survey Explorer \citep[WISE;][]{Wright_2010}, Spitzer Space Telescope \citep{Werner_2004} and SPHEREx \citep{SPHEREx_SPIE}, looking for extremely cold companions to nearby stars and brown dwarfs with the James Webb Space Telescope \citep[JWST; e.g.,][]{Calissendorff_2023,De_Furio_2025,Albert_2025}, and wide-area searches with NEO Surveyor \citep{NEO_Surveyor,Kirkpatrick_NEOCam} and Roman Space Telescope \citep{WFIRST} in the future.

Additionally, JWST spectroscopic and photometric observations to date have revealed remarkable diversity among the population of nearby Y dwarfs \citep[e.g.,][]{Faherty_2021,Beiler_sample,Albert_2025,Leggett_2025,Leggett_2026}. Examples include the discovery of methane emission from CWISEP J193518.59-154620.3 \citep{Marocco_2019,Faherty_methane} and the detection of silane in the atmosphere of halo T/Y dwarf WISEA J153429.75-104303.3 \citep{Faherty_W1534}. Beyond completing the nearby Y dwarf census, each $T_{\rm eff} \lesssim 500$~K brown dwarf discovered presents a new opportunity to probe the broad range of behaviors and abundances now measurable with JWST spectroscopy. A large sample of well-characterized Y dwarfs is needed, as an array of factors are expected to contribute to their differing observational properties. These factors include metallicity \citep[e.g.,][]{Meisner_2021,Burgasser_PH3}, gravity \citep[e.g.,][]{Robbins_2023}, age \citep[e.g.,][]{Leggett_2017}, binarity \citep[e.g.,][]{Calissendorff_2023,De_Furio_2025}, clouds \citep[e.g.,][]{Morley_2012,Morley_2014}, and disequilibrium chemistry \citep[e.g.,][]{Leggett_2023,Rowland_w0855}.

Here we report the discovery of \objname, a high proper motion object found using WISE data, with a serendipitous Spitzer ch1$-$ch2 color limit\footnote{The Spitzer/IRAC ch1 (ch2) band with central wavelength $\approx$3.6~$\mu$m ($\approx$4.5~$\mu$m) is also sometimes referred to as [3.6] ([4.5]) in the literature.} suggesting that it is likely a Y-type brown dwarf within 10~pc of the Sun. In $\S$\ref{sec:discovery} we discuss our search for faint, fast-moving objects with the SMDET neural network architecture \citep{SMDET_paper} and the resultant discovery of \objname. In $\S$\ref{sec:characterization} we characterize \objname~to the extent currently possible using archival and follow-up data. In $\S$\ref{sec:discussion} we contextualize \objname~within the presently known sample of cold, nearby brown dwarfs. We conclude in $\S$\ref{sec:conclusion}.

\section{Search and Discovery} \label{sec:discovery}

\subsection{Time-resolved unWISE Coadds} \label{sec:coadds}

Time-resolved unWISE coadds \citep{tr_coadds} stack individual WISE exposures within $\sim$1 day time intervals (``sky passes''), which occur roughly every six months at a given sky location based on the WISE scanning pattern. This enables the detection of fainter solar neighborhood proper motion objects than would be possible with individual WISE exposures, while incurring essentially no loss of information. For instance, the highest known proper motion of any star or brown dwarf is $\sim 10\farcs4$/yr (Barnard's Star; \citealt{Barnard_1916}), which corresponds to a negligible angular translation over the course of $\sim$1 day relative to the WISE 3-5~$\mu$m full width at half maximum (FWHM) of $\sim$6$''$.

We used all-sky time-resolved unWISE coadds spanning 2010-2020 as input data for the high proper motion object search through which we discovered \objname~(see $\S$\ref{sec:smdet}). During the time period after our full-sky search was conducted, additional unWISE time-resolved coadd epochs corresponding to the most recent WISE sky passes through the end of the NEOWISE mission \citep[mid-2024;][]{Mainzer_2011} have become available, allowing us to further confirm the motion of \objname~beyond the timespan of the original discovery images.

\subsection{SMDET} \label{sec:smdet}

The SMDET neural network methodology, including a diagram of SMDET's architecture, is presented in \cite{SMDET_paper}. SMDET employs a recurrent convolutional neural network to analyze time-series astronomical images, specifically looking for faint, fast-moving objects (FFOs; with large motion being a proxy for nearness in the context of brown dwarf searches). This neural network was trained using synthetic FFOs injected into unWISE image data, and the SMDET analysis subdivides the sky spatially into a set of 2.9$' \times 2.9'$ image sequences. The SMDET network architecture, which incorporates 3D convolutional layers and 2D convolutional long short-term memory layers, generates outputs that include segmented image sequences and reproductions of any FFOs it identifies. These outputs are used to rank the image sequences based on the probability of containing an FFO and to pinpoint the exact location of each candidate FFO.

The input data for the SMDET deployment that discovered \objname~consisted of 16 time-resolved unWISE coadd images per sky location for each of the two bluest WISE channels (W1 $\approx 3.4$~$\mu$m and W2 $\approx 4.6$~$\mu$m), covering a period of $\approx 10.5$ years at each sky location (early 2010 to late 2020). This SMDET run was performed several years prior to the end of the NEOWISE mission, and so did not span the full $\approx 14.5$ year WISE/NEOWISE time baseline now available. We deployed SMDET across the entire sky, and the top 10,000 highest-ranked image sequences, centered on SMDET's FFO coordinates, were then visually inspected by D. Caselden, resulting in the initial discovery of many candidate high proper motion objects including \objname. Other SMDET moving object discoveries have previously been presented in \cite{Brooks_2022} and \cite{Brooks_SMDET_sample}, with additional SMDET discoveries to be presented in Raghu et al. (in prep.).

\subsection{\objname: Discovery \& High-significance Confirmation with Spitzer}

We initially discovered \objname~based solely on W1 and W2 time-resolved unWISE coadds, analyzed with SMDET. However, as \objname~is located near the Galactic plane ($l_{gal}$, $b_{gal} \approx$ 63.6$^{\circ}$, +2.25$^{\circ}$) its sky position has serendipitous Spitzer IRAC \citep{Fazio_2004} ch1 and ch2 imaging from the Deep GLIMPSE survey \citep{Deep_GLIMPSE_AAS,GLIMPSE_surveys,deep_GLIMPSE_ARCHIVE}, acquired in late 2012 December. \objname~is present as a ch2-only source in the Deep GLIMPSE catalog, with ch2 = $14.574 \pm 0.089$ mag (Vega) and a non-detection in ch1. The Deep GLIMPSE counterpart's designation is SSTGLMDPA G063.5983+02.2549. The Spitzer confirmation of \objname~is vital because Spitzer/IRAC (FWHM $\approx 2''$) provides $\gtrsim 3\times$ better angular resolution than WISE, allowing Spitzer to resolve \objname~and accurately measure its 4.5~$\mu$m flux with high signal-to-noise.

\begin{figure}[!ht]
    \centering
\includegraphics[width=1.0\textwidth]{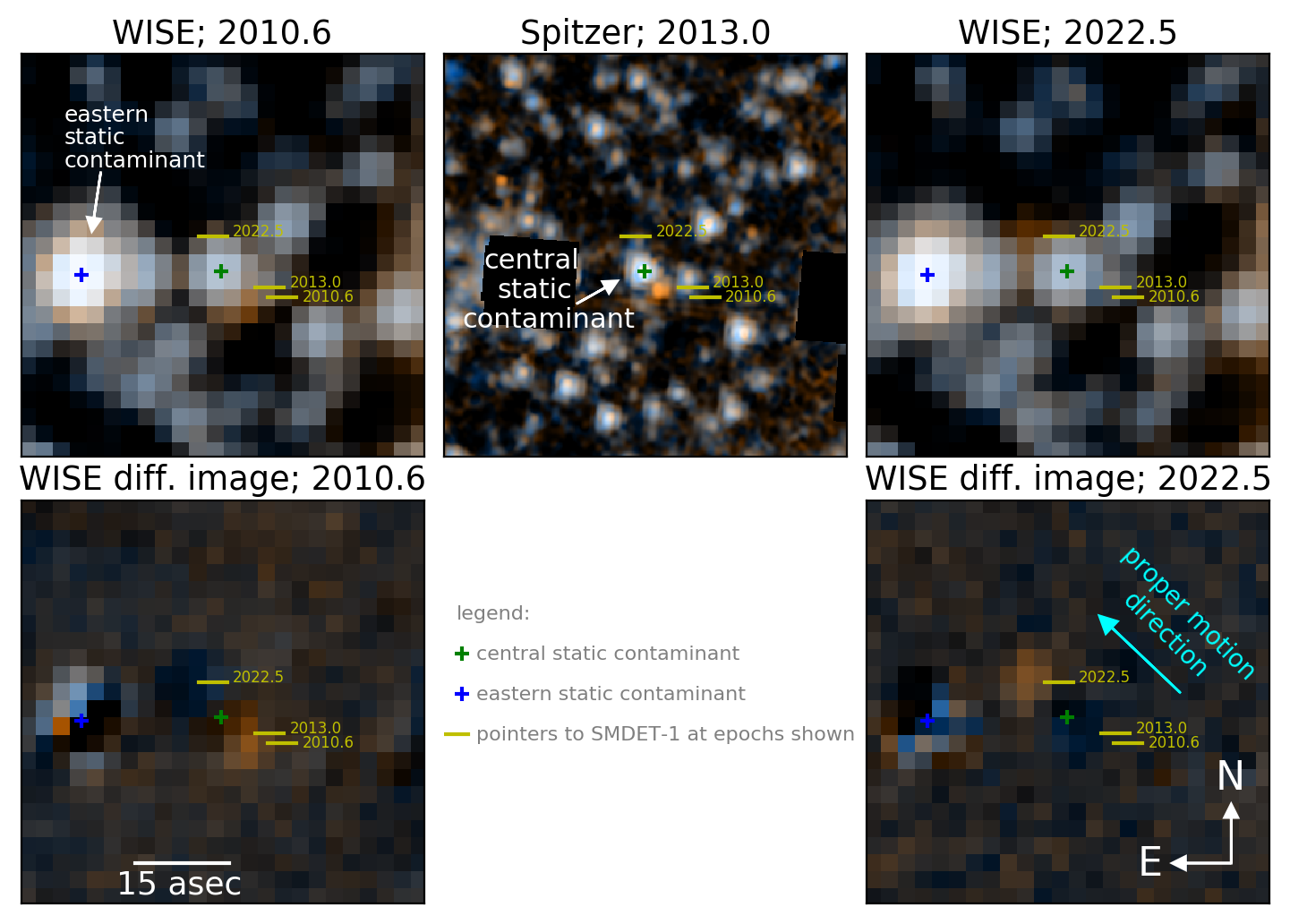}
    \vskip -1em
    \caption{Top: time-series unWISE and Spitzer cutouts with no template subtraction applied, showing \objname's northeasterly motion over $\approx$~12 years. Each panel displays a two-color composite where $\lambda \approx 3.5$~$\mu$m is the blue channel and $\lambda \approx 4.5$~$\mu$m is the red channel. Main sequence stars appear blue-white, whereas a very cold brown dwarf or highly reddened background source will appear deep orange. \objname~is resolved as an orange point source to the central contaminant's southwest by Spitzer at epoch 2013.0 (top center), and the orange WISE counterpart is seen at a qualitatively similar location in 2010, though blended (top left). By 2022.5, \objname~has  `crossed over' to the northeast of the central static contaminant. Bottom: WISE difference images confirm this motion. A horizontal yellow line points to \objname~at each epoch shown, with \objname~being near the left end of each horizontal yellow line at the relevant epoch. Each yellow line is labeled in each panel with its corresponding epoch. The blue and green plus marks indicate two static, relatively bright background contaminants (the brighter WISE contaminant labeled with a blue plus mark is resolved by 2MASS as a pair of sources). The bottom center panel displays no image because there is only one Spitzer epoch available and hence no Spitzer difference image can be created.}\label{fig:images}
\end{figure}

Figure \ref{fig:images} shows \objname's northeasterly motion within a subset of the available WISE epochs, WISE difference images, and Deep GLIMPSE. During the WISE/NEOWISE mission timespan (early 2010 to mid 2024), \objname~passes in front of a brighter background contaminant (labeled with a green plus mark in Figure \ref{fig:images}), likely explaining why \objname~was not recognized until now. At early WISE epochs and in the Deep GLIMPSE imaging, \objname~lies slightly to the southwest of this contaminant, appearing blended in WISE but resolved in Spitzer. By the end of the NEOWISE mission, \objname~has `crossed over' to the northeast of this background contaminant though still remains blended with this contaminant and is best visually ascertained in WISE difference images (bottom row of Figure \ref{fig:images}). The WISE image at top left in Figure \ref{fig:images} is an unWISE coadd spanning 2010.3-2010.8, with mean epoch 2010.6. The WISE image at top right in Figure \ref{fig:images} is an unWISE coadd spanning 2022.3-2022.8, with mean epoch 2022.5. The epoch 2010.6 WISE difference image (lower left of Figure \ref{fig:images}) was created by subtracting off a template consisting of all 2020.3-2022.8 WISE data from the epoch 2010.6 WISE image shown at upper left. The epoch 2022.5 difference image (lower right of Figure \ref{fig:images}) was created by subtracting off a template consisting of all 2010.3-2015.8 WISE data from the epoch 2022.5 WISE image shown at upper right. 

At the end of the NEOWISE mission, \objname~is additionally blended (at WISE's angular resolution) with an even brighter contaminant to the east (labeled with a blue plus mark in Figure \ref{fig:images}; 2MASS resolves this brighter contaminant into two separate sources). In the Figure \ref{fig:images} WISE images with epoch = 2022.5, there is no longer a WISE counterpart at/near the location of the ch2-only Deep GLIMPSE catalog source, as expected for an object moving on order 1$''$/yr. Although Figure \ref{fig:images} only shows two WISE epochs, \objname~is visible in additional epochs of WISE difference images in a manner consistent with the linear motion trajectory illustrated in Figure \ref{fig:images} and a constant W2 flux.

\section{Further Investigation of \objname} \label{sec:characterization}

\subsection{Spitzer Color Limit}
\label{sec:color_lim}

We sought to translate the Deep GLIMPSE catalog's ch1 non-detection of \objname~into a ch1 magnitude lower limit. To do so, we retrieved all $\sim 1,600$ Deep GLIMPSE objects within 176$''$ of the \objname~ch2 counterpart. This radius was chosen because it corresponds to an area equivalent to the IRAC per-channel field of view. We then fit a second order polynomial to the trend of ch1 signal-to-noise (S/N) versus ch1 magnitude. We find that this polynomial reaches S/N = 5 at ch1 = 17.38 mag (Vega), which we therefore adopt as our 5$\sigma$ ch1 limit for \objname. This 5$\sigma$ ch1 limit corresponds to a Spitzer color limit of ch1-ch2 $>$ 2.81 mag.

We note that the Deep GLIMPSE Data Description's stated survey-wide ch1 sensitivity limit\footnote{\url{https://irsa.ipac.caltech.edu/data/SPITZER/GLIMPSE/doc/deepglimpse_dataprod_v1.3.pdf}, Table 2.} is ch1 = 17.80 mag (Vega), 0.42~mag fainter than our adopted 5$\sigma$ ch1 depth at the location of \objname; our adopted ch1 depth is more conservative. The Deep GLIMPSE Data Description does not state precisely how its quoted ch1 = 17.80~mag sensitivity limit is defined, but if this were a 3$\sigma$ limit, then it would agree with our derived 5$\sigma$ limit to within $< 0.15$~mag. The Deep GLIMPSE catalog nulls out ch1 fluxes below S/N = 5, so in any case we find it most appropriate to employ a 5$\sigma$ limit rather than a 3$\sigma$ limit.

\subsection{Spitzer ch1 Forced Photometry}
\label{sec:forcedphot}

Given the ch1 non-detection of \objname~reported by Deep GLIMPSE, we performed forced photometry on the relevant ch1 imaging using the ch2 position of \objname~measured at the same epoch. Forced photometry has the potential to further quantify any sub-threshold ch1 flux which may be present at \objname's location despite the lack of a 5$\sigma$ ch1 detection. We began by building a custom mosaic of the Deep GLIMPSE ch1 observations using the MOPEX software \citep{Makavoz_2005}. We then derived a best-fit ch1 point source amplitude at the sky location of the ch2 \objname~detection after subtracting neighboring contaminants.

\begin{figure}[!ht]
    \centering
    \includegraphics{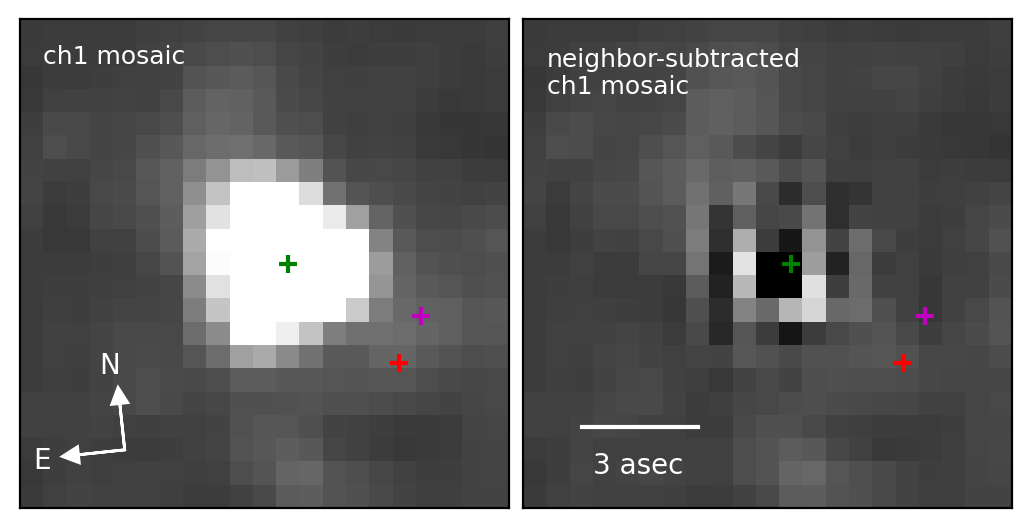}
    \caption{Left: Spitzer ch1 mosaic built from late-2012 Deep GLIMPSE observations. As in Figure \ref{fig:images}, the central static contaminant is labeled with a green plus mark. The magenta plus mark denotes the location of another much fainter background contaminant. The red plus mark indicates the location of the Spitzer ch2 detection of \objname~at the same epoch as this ch1 imaging. Right: ch1 mosaic after subtracting the central static contaminant and the fainter western contaminant. There appears to be a very small amount of remaining ch1 flux near/at the position of \objname~though this residual flux may not be pointlike, might be due at least in part to imperfect removal of the bright central contaminant's PSF wings, and has a significance of only 3.4$\sigma$ when fit with a point source at \objname's location. These images illustrate that there may be a modicum of ch1 flux at \objname's position after accounting for nearby contaminants, but there is not a definitive ch1 detection.}\label{fig:forcedphot}
\end{figure}

Figure \ref{fig:forcedphot} shows a close-up of the ch1 imaging data near \objname~and illustrates our forced photometry procedure. We built a pixelized point spread function (PSF) model using $\approx 60$ bright, nearby stars in the ch1 mosaic. We then subtracted the bright, central contaminant (green plus mark in Figure \ref{fig:forcedphot}) by scaling the PSF to match the measured flux of this contaminant in a 3 pixel radius aperture and shifting the PSF to account for the contaminant's subpixel centroid location. The much fainter western contaminant (magenta plus mark) is not detected at 5$\sigma$ in Deep GLIMPSE, but is detected by both UKIDSS \citep{UKIDSS} and Gaia \citep[Gaia DR3 2031188170012798208;][]{gaia_mission,Gaia_DR3}. The UKIDSS JHK colors of the western contaminant are virtually identical to those of the central contaminant, so we assigned the western contaminant a ch1 mag by assuming it has the same K-ch1 color as the central contaminant. Using this adopted ch1 mag, we subtracted an appropriately scaled/shifted PSF model at the western contaminant's location. The resulting `neighbor-subtracted' ch1 mosaic after removing these two contaminants is shown in the right panel of Figure \ref{fig:forcedphot}. Visually, there does appear to be a very small amount of residual ch1 flux near/at the location of \objname, though this residual flux may not be pointlike and might be at least partially attributable to imperfect PSF modeling of the far brighter central contaminant.

We derived a ch1 forced photometry flux for \objname~by fitting a point source to the neighbor-subtracted ch1 image at the  \objname~ch2 position (red plus mark in Figure \ref{fig:forcedphot}). We find a sub-threshold (3.4$\sigma$ significance) ch1 flux corresponding to ch1 = 17.66$^{+0.38}_{-0.28}$ mag (Vega), consistent with the 5$\sigma$ ch1 $>$ 17.38 mag (Vega) limit we derived in $\S$\ref{sec:color_lim}. We favor the $\S$\ref{sec:color_lim} ch1 limit over the possible sub-threshold ch1 `detection', and adopt the ch1 limit throughout the remainder of this work for the following reasons: (1) the possible ch1 counterpart's flux is less than 5$\sigma$ significant, (2) the residual ch1 flux near \objname's predicted position does not definitively appear pointlike and does not align optimally with the predicted centroid, and (3) this residual ch1 flux may be at least in part due to imperfect modeling of the $> 100\times$ brighter central contaminant.

\subsection{unTimely Catalog and \texttt{crowdsource}} \label{sec:crowdsource}

We consulted the unTimely Catalog \citep{Meisner_2022,ucx} of sources extracted from time-resolved unWISE coadds to search for any WISE detections of \objname. \objname~has the best chance of being detected in unTimely during the very early WISE mission (2010-2011; when \objname~lies somewhat to the southwest of its main contaminant) and near the end of the NEOWISE mission (2022-2024; when \objname~lies somewhat to the northeast of its main contaminant). unTimely provides two detections of \objname, corresponding to the two WISE sky passes covering the relevant sky region during the pre-hibernation WISE mission. These unTimely detections have W2 magnitudes that match the late-2012 Spitzer ch2 mag  well considering the severe blending in WISE (W2 = 14.59 $\pm$ 0.055 mag, 14.70 $\pm$ 0.057 mag; note that for cold brown dwarfs we expect ch2 and W2 magnitudes to be essentially the same per Figure 15 of \citealt{Kirkpatrick_2021a}). During each of these two pre-hibernation WISE sky passes, there is no W1 unTimely detection within 2$''$ of the W2 \objname~counterpart. The distribution of W1 unTimely magnitudes near \objname~in these two epochs is sharply peaked around W1 = 15 mag (Vega), falling off rapidly toward fainter magnitudes. Hence we quote a limit of W1 $>$ 15 mag (Vega) for \objname~in Table \ref{tab:data}.

The unTimely Catalog only includes epochs up to approximately the end of calendar 2020, containing no post-reactivation WISE detections of \objname. To obtain unTimely-like deep epochal WISE catalogs through the end of the NEOWISE mission, we ran the same \verb|crowdsource| code \citep{decaps,unWISE_catalog} used to generate unTimely on all remaining unWISE time-resolved coadds covering \objname's position through the end of the NEOWISE mission. We thereby obtained one additional unWISE detection of \objname, at epoch $\approx$ 2022.8. Combining  our  \verb|crowdsource| detection of \objname~with its two unTimely Catalog detections and the Deep GLIMPSE catalog's ch2 counterpart, we have 4 astrometric epochs available, spanning $\approx$ 12.4 years. Table \ref{tab:data} provides the proper motion solution resulting from our fit of these four WISE and Spitzer detections (the MJD listed in Table \ref{tab:data} is the reference epoch at which~\objname's RA and Dec are quoted in Table \ref{tab:data}). The formal uncertainties on $\mu_{\alpha}$, $\mu_{\delta}$ quoted in Table \ref{tab:data} may be underestimated due to \objname's severe blending in WISE. The reduced chi-squared for the $\mu_{\delta}$ fit is quite low ($\chi^2_{\nu} = 0.65$), though the reduced chi-squared for $\mu_{\alpha}$ is poor ($\chi^2_{\nu} = 13.9$). We did not attempt to fit a parallax for \objname~given that there are only 4 epochs of astrometry, 3 of which are heavily blended WISE detections. Additional astrometry of \objname~based on higher angular resolution imaging is needed.

\subsection{Palomar Follow-up Imaging}
\label{sec:palomar}

We obtained deep J-band follow-up imaging (PI: Marocco) of the sky region containing \objname~on 2023 April 30 (UTC) using the Wide Field Infrared Camera \citep[WIRC;][]{WIRC} at the Palomar Hale 200-inch telescope. We used a 15-point dithering with 2 minutes per dither ($4 \times 30$~s exposures). The delivered image quality was 1$''$ FWHM, and we determine a 5$\sigma$ point source depth of J = 21.16 mag (Vega) for the reduced, coadded mosaic by cross-matching against UKIDSS \citep{UKIDSS} photometry. There is no Palomar/WIRC counterpart within the 1$\sigma$ positional uncertainty ellipse of \objname~at the relevant epoch according to our Table \ref{tab:data} proper motion solution, allowing us to place a 5$\sigma$ limit of J $>$ 21.16 mag, which we report in Table \ref{tab:data}.

\subsection{UKIDSS Galactic Plane Survey Archival Near Infrared Imaging} \label{sec:ukidss}

The UKIDSS Galactic Plane Survey \citep{UKIDSS_GPS} imaged the sky location of \objname~in the J, H, and K bands on 2010 August 30, then again in K band on 2013 May 17. There is no J, H, or K counterpart within $1''$ of \objname's predicted position at the UKIDSS 2010 epoch. The closest UKIDSS source is 1\farcs4 to the north north west, appears at the same sky location in the 2013 UKIDSS image, and has well-measured Gaia proper motions smaller than 5 mas/yr in each component. We therefore conclude that this UKIDSS/Gaia source is not a potential counterpart to \objname. The 2010 UKIDSS epoch is bracketed by two unTimely Catalog detections of \objname~($\S$\ref{sec:crowdsource}), both within 4 months of the 2010 UKIDSS epoch, such that our predicted \objname~position at this epoch is well-constrained.

At the 2013 UKIDSS epoch, there is a K-band source $\approx 0\farcs6$ from \objname's predicted position, though this K-band source appears at the same location in 2010 UKIDSS imaging and has well-measured Gaia proper motions smaller than 10 mas/yr in each component. We therefore conclude that this UKIDSS/Gaia source is not a potential counterpart to \objname. No other UKIDSS source is within 2\farcs7 of \objname's predicted position in the 2013 UKIDSS K-band image. Finding no UKIDSS counterpart, we apply the methodology from \cite{Schneider_2020} to derive 5$\sigma$ limits on the JHK magnitudes of \objname. These H-band and K-band limits are listed in Table \ref{tab:data}, whereas Table \ref{tab:data} reports our WIRC J-band limit, which is deeper than the J $>$ 19.82 mag 5$\sigma$ limit from UKIDSS. 

\subsection{Early SPHEREx Data} \label{sec:spherex}

We do not expect SPHEREx to detect \objname~in the near-infrared, as J $>$ 21.16 mag (Vega) is several magnitudes fainter than the SPHEREx all-sky sensitivity in this wavelength range \citep{Bock_SPHEREx}. On the other hand, the nominal 1-year all-sky SPHEREx 4-5~$\mu$m depth (not accounting for the adverse effect of confusion in highly crowded fields) corresponds to $\sim$ 14.5 mag Vega, similar to the ch2 brightness of \objname, suggesting that there may be a chance of obtaining a weak SPHEREx detection at 4-5~$\mu$m.

\objname's sky location was observed by SPHEREx \citep{SPHEREx_Dore,SPHEREx_SPIE,spherex_qr2_2025} in 139 linear variable filter (LVF) images between 2025 April 24 and 2025 November 1 (UTC). These images are split into two sky passes, the first with mean MJD = 60798 and the second with mean MJD = 60949. We visually examined cutouts of all of these SPHEREx LVF images in the vicinity of \objname. Given SPHEREx's substantially larger pixel size compared to WISE (6\farcs2 versus 2\farcs75), strong crowding, and the relatively limited signal-to-noise available in individual LVF images, there is not conclusive visual evidence of an \objname~counterpart or lack thereof at \objname's predicted location.

We used the ‘SPHEREx Photometry and Image Fitting Framework’ \citep[SPIFF;][]{SPIFF,SPIFF_zenodo} to attempt to extract a SPHEREx spectrum of \objname. We were unable to detect \objname~with SPIFF, either through a standard spectrophotometric extraction, or through alternative means such as stacking SPHEREx images with similar wavelengths near \objname's position and/or attempting to subtract off neighboring contaminants. The undersampling of SPHEREx images makes such attempts at coaddition and image differencing particularly challenging. We also attempted to obtain early SPHEREx spectrophotometry of \objname~using the IRSA/SPHEREx Spectrophotometry Tool, but found the results inconclusive. Appendix \ref{sec:appendix} provides further details and illustrates the inconclusive nature of the present SPHEREx data.

Using the first $\sim 1$ year of SPHEREx data, \cite{SPIFF} found that ultracool dwarfs with W2 brighter than 14 mag (Vega) generally had high quality (average S/N per spectral channel $\gtrsim 10$) SPHEREx spectra. \objname~is $\approx 0.6$ mag fainter than this approximate W2 $\approx$ 14~mag threshold appropriate for a typical uncrowded sky region. Based on SPHEREx's depth \citep{Bock_SPHEREx}, we expect SPHEREx to be most sensitive to a Y dwarf at $\lambda \approx 4$-5~$\mu$m. WISE W2 catalogs near \objname~show that at $\sim 6''$ resolution (similar to that of SPHEREx), this sky region's crowding effectively reduces sensitivity (compared to typical uncrowded fields) by $\approx 1.5$ magnitudes, meaning that we might expect a Y dwarf near \objname's sky location to require W2 $\lesssim$ 12.5 mag (Vega) to yield a high-quality first-year SPHEREx spectrum. This crowding-adjusted SPHEREx threshold is $\sim 2$~magnitudes brighter than \objname, hence our lack of a conclusive detection of this object in early SPHEREx data is reasonable.

\begin{deluxetable}{lccc}
\tablecaption{Compiled Data for \objname \label{tab:data}}
\tablehead{
\colhead{Parameter} & \colhead{Value} & \colhead{Ref.}}
\startdata
\cutinhead{Astrometric Fit}
$\mu$$_{\alpha}$ (mas yr$^{-1}$) & 1059.5 $\pm$ 45.4 & 1\\
$\mu$$_{\delta}$ (mas yr$^{-1}$)  & 714.3 $\pm$ 46.3 & 1\\
RA & 295.791939 deg $\pm$ 149 mas & 1 \\
Dec & +28.132809 deg $\pm$ 152 mas & 1 \\
MJD (d) & 55972.68 & 1 \\
\cutinhead{Photometry}
J$_{MKO}$ (mag; Vega) & $> 21.16$ (5$\sigma$) & 1 \\
H$_{MKO}$ (mag; Vega) & $> 19.09$ (5$\sigma$) & 1 \\
K$_{MKO}$ (mag; Vega) & $> 18.37$ (5$\sigma$) & 1 \\
W1 (mag; Vega) & $>$ 15 & 1\\
W2  (mag; Vega) & 14.65 $\pm$ 0.06 & 1\\
ch1 (mag; Vega) & $> 17.38$ (5$\sigma$) & 1 \\
ch2  (mag; Vega) & 14.574 $\pm$ 0.089 & 2 \\
\cutinhead{Inferred Properties}
$T_{\rm eff}$ (K)  & $<$ 391 & 1\\
$M_{\rm ch2}$ (mag)  & $> 15.22$ & 1\\
$d$ (pc)  & $< 7.4$ & 1\\
$v_{tan}$ (km/s)  & $< 44.9$ & 1\\
\enddata
\tablerefs{ (1) This work; (2) \citealt{GLIMPSE_surveys}. }
\end{deluxetable}

\section{Discussion} \label{sec:discussion}

\subsection{\objname~in Context}

\begin{figure} [ht]
    \centering
    \includegraphics[width=0.8\textwidth]{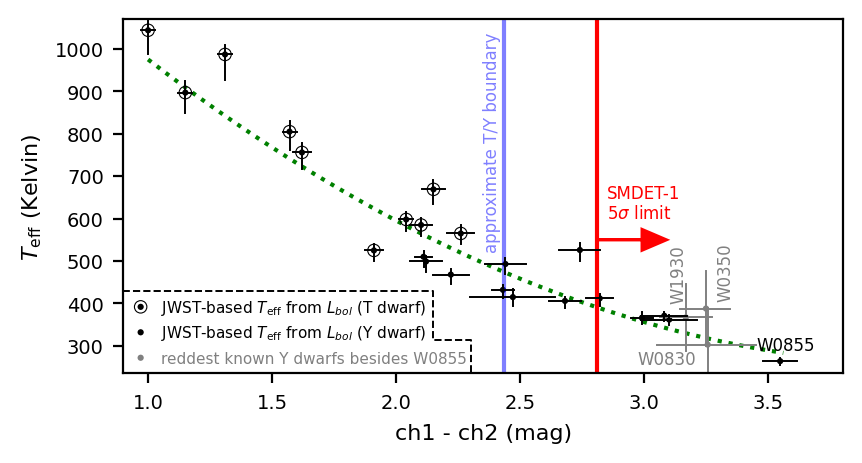}
    \vskip -1em
    \caption{Spitzer ch1-ch2 colors of $T_{\rm eff} \lesssim 1000$ K brown dwarfs correlate strongly with their temperatures, with redder ch1-ch2 color generally indicating colder temperature. Black data points are JWST-based temperature measurements from bolometric luminosity \citep{Beiler_sample}. Gray data points with large gray error bars are cases that lack JWST-based bolometric luminosities, and have uncertain photometric temperatures estimated based on a \cite{Kirkpatrick_2021a} polynomial relation. WISEA J083011.95+283716.0 \citep{Bardalez_2020} is labeled W0830, WISE J035000.32-565830.2 \citep{Kirkpatrick_2012} is labeled W0350, and WISEA J193054.55-205949.4 \citep{Meisner_2020b} is labeled W1930.  WISE 0855$-$0714 \citep{Luhman_2014} is labeled W0855, and we adopt its JWST-based temperature estimate from \cite{Rowland_w0855}. The approximate T/Y boundary in terms of ch1-ch2 color is denoted by a blue vertical line with abscissa value calculated according to the ch1-ch2(SpT) polynomial relation from \cite{Kirkpatrick_2021a}. Our \objname~Deep GLIMPSE ch1-ch2 color limit ($\S$\ref{sec:color_lim}; red vertical line with red arrow) corresponds to a Y dwarf Spitzer phototype. The dotted green line is the polynomial we have fit to the black data points.}\label{fig:color_trend}
\end{figure}

Employing an absolute magnitude $M_{\rm ch2}$ versus ch1$-$ch2 polynomial relation \citep{Kirkpatrick_2021a}, \objname's ch1$-$ch2 color lower limit of 2.81 mag implies a very nearby distance $d < 7.4$~pc. The $M_{\rm ch2} > 15.22$~mag limit based on ch1$-$ch2 color represents a stronger constraint than $M_{\rm ch2} > 15.0$~mag derived from J-ch2 $ > 6.59$~mag \citep{Kirkpatrick_2021a}. \cite{Kirkpatrick_2021a} do not provide a polynomial relation for $T_{\rm eff}$ as a function of ch1$-$ch2. To enable an initial temperature limit, we therefore fit a second order polynomial to the JWST-based effective temperatures \citep{Beiler_sample} shown in Figure \ref{fig:color_trend}, which leverage high-quality bolometric luminosity measurements. For WISE 0855$-$0714 \citep[W0855;][]{Luhman_2014}, we adopt a temperature of 264 $\pm$ 8~K from \cite{Rowland_w0855}, which also uses the \cite{Beiler_sample} methodology. From our second order $T_{\rm eff}$(ch1$-$ch2) polynomial, we find that \objname's ch1$-$ch2 $>$ 2.81 mag color corresponds to a photometric temperature limit of $T_{\rm eff} < $ 391~K. \cite{Leggett_2026} also provides polynomial relations for $T_{\rm eff}$ applicable in the relevant brown dwarf temperature regime, including as a function of ch1$-$ch2, J-ch2, and $M_{\rm ch2}$. These three \cite{Leggett_2026} polynomial relations yield limits of $T_{\rm eff} < 383$~K, $T_{\rm eff} < 397$~K, and $T_{\rm eff} < 395$~K, respectively, consistent with the limit from our polynomial fit. We note that recent studies have shown that metal-poor objects can be very large outliers relative to the general trend of brown dwarf temperature versus ch1$-$ch2 color \citep[e.g.,][]{Faherty_W1534}.

Figure \ref{fig:color_trend} shows that \objname's ch1-ch2 color limit places it among Y dwarfs rather than T dwarfs. Indeed, the (approximate) T/Y boundary lies at ch1-ch2 = 2.44 mag \citep{Kirkpatrick_2021a}, substantially bluer than our \objname~ch1-ch2 color limit. Our temperature range $T_{\rm eff} < $ 391~K is also characteristic of a Y dwarf rather than a T dwarf. We therefore consider \objname~to have a Y dwarf Spitzer phototype, and applying the \citealt{Kirkpatrick_2021a} SpT(ch1-ch2) polynomial relation gives a phototype limit of $>$~Y0.8. The \citealt{Kirkpatrick_2021a} SpT(J-ch2) polynomial relation gives a weaker phototype limit of $>$~Y0.1.

The JHK non-detections of $\S$\ref{sec:palomar}-\ref{sec:ukidss} (see Table \ref{tab:data}) are consistent with a Y dwarf spectral type for \objname~but constrain its temperature less stringently than does its ch1-ch2 color limit. The Palomar-Spitzer color limit of J-ch2 $>$ 6.59 mag is $\approx 1$ magnitude redder than the approximate T/Y boundary in J-ch2 color \citep{Kirkpatrick_2021a}.

Adopting our $d < 7.4$ pc distance limit, the total proper motion of \objname~implies a tangential velocity $v_{tan} < 44.9$ km/s. This $v_{tan}$ limit is consistent with the median tangential velocity of 30.8 km/s for LTY dwarfs within 20 pc \citep{Kirkpatrick_2021a}, a population dominated by the thin disk \citep{Dino_SMART}. In the absence of a trigonometric parallax, reduced proper motion \citep{Jones_1972} is a proxy for gauging an object's combination of kinematics and luminosity (e.g., large reduced proper motion points to high $v_{tan}$ and/or low luminosity). The ch2 reduced proper motion of \objname, $H_{\rm ch2} \approx 20.1$ mag, is unremarkable with respect to the distribution of $H_{\rm ch2}$ among known Y dwarfs (e.g., Figure 5 of \citealt{Meisner_2023}).

\subsection{Implications of Sub-threshold ch1 Forced Photometry Flux}

If one were to adopt the sub-threshold forced photometry ch1 flux of $\S$\ref{sec:forcedphot} for \objname, this would enable alternative estimates of \objname's properties. The corresponding ch1-ch2 color would be 3.09$^{+0.39}_{-0.29}$ mag. The photometric distance estimate would be 5.6$^{+2.3}_{-1.8}$~pc, derived using the \cite{Kirkpatrick_2021a} $M_{\rm ch2}$(ch1-ch2) polynomial relation. The photometric temperature estimate would be $T_{\rm eff} = 349^{+39}_{-59}$~K using the \cite{Leggett_2026} $T_{\rm eff}$(ch1-ch2) polynomial. \objname's predicted J magnitude would be 22.6$^{+1.8}_{-1.3}$ mag (Vega) using the \cite{Kirkpatrick_2021a} J-ch2(ch1-ch2) polynomial. The ch1-ch2 = 3.09$^{+0.39}_{-0.29}$ mag color is too red to employ the \cite{Kirkpatrick_2021a} SpT(ch1-ch2) photometric type polynomial relation, which is not applicable at ch1-ch2 $>$ 3 mag.

\subsection{Future Potentially Relevant Survey Data}

As SPHEREx accumulates additional data (and as SPHEREx tooling optimized for the highly crowded, low S/N regime matures), it will be of interest to continue checking SPHEREx at the relevant locations of \objname. \objname~may not benefit from NEO Surveyor data, as its sky location corresponds to ecliptic coordinates ($\lambda$, $\beta$) $\approx$ (305.4$^{\circ}$, 48.5$^{\circ}$) whereas NEO Surveyor is expected to cover only $|\beta| < 40^{\circ}$ at baseline. NEO Surveyor will also have angular resolution similar to that of WISE W1/W2, such that crowding would still present difficulties even if that mission were to image the location of \objname. Meanwhile Euclid \citep{Euclid_mission} will not survey the Galactic plane. Roman may image \objname's sky location in the near infrared depending on the exact footprint chosen for its anticipated Galactic plane survey \citep{GRIPS,Roman_plane_committee}. \objname~is too far north given its right ascension to be observed via the LSST \citep{Ivezic_LSST} Wide-Fast-Deep survey\footnote{\url{https://survey-strategy.lsst.io/baseline/index.html}}.

\section{Conclusion \& Outlook} \label{sec:conclusion}

We have presented the discovery of \objname, a candidate nearby Y dwarf with a photometrically estimated distance of less than 10~pc. Thus far, \objname~is securely detected only at 4-5~$\mu$m by Spitzer (ch2) and WISE (W2). If \objname~has a $T_{\rm eff}$ value close to its photometrically estimated temperature upper limit of 391~K, it may be detectable at J-band via dedicated follow-up with large ground-based telescopes. \objname's ch1-ch2 color lower limit corresponds to a J-ch2 color lower limit of 6.85 mag according to the relevant \cite{Kirkpatrick_2021a} polynomial relation. This yields an expectation of J $>$ 21.4 mag (Vega) for \objname. The relatively bright end of this J magnitude range is within reach of e.g., Gemini Observatory \citep[e.g.,][]{Leggett_2014_phot}. If \objname~is substantially cooler than our photometrically estimated temperature upper limit, then obtaining any future detections of it may require Hubble Space Telescope, Roman, or JWST follow-up.

Measuring a trigonometric parallax for \objname~will be particularly important. Given the significant blending that affects its WISE detections, we expect that obtaining a high-S/N trigonometric parallax for \objname~would require combining its Spitzer ch2 detection with (at least) two additional epochs of astrometry from observations with FWHM~$\lesssim 1''$, ideally sampling opposite ends of the parallactic ellipse. Dedicated follow-up observations with Hubble's Wide Field Camera 3 \citep[e.g.,][]{Marocco_2026}, Roman's Wide Field Instrument, or JWST/NIRCam \citep{NIRCam_performance} would be capable of providing such astrometry.

At any rate, \objname~serves to emphasize that there is still remaining discovery space for very cold and close brown dwarfs within the WISE and Spitzer archives, particularly when these are paired with novel pixel-level analysis methodologies. The discovery of \objname~suggests that deep learning may be a promising avenue for identifying faint moving objects in crowded fields when applied to archival and future survey datasets.

\begin{acknowledgments}
We thank the anonymous referee for their suggestions which have resulted in an improved manuscript. We thank S.~K. Leggett for valuable comments on this manuscript. The work of AMM is supported by NOIRLab, which is managed by the Association of Universities for Research in Astronomy (AURA) under a cooperative agreement with the U.S. National Science Foundation. Support for US investigators in programs 3558 and 6084 was provided by NASA through a grant from the Space Telescope Science Institute, which is operated by the Association of Universities for Research in Astronomy, Inc., under NASA contract NAS 5-03127.

Based on observations obtained at the Hale Telescope, Palomar Observatory, as part of a collaborative agreement between the Caltech Optical Observatories and the Jet Propulsion Laboratory [operated by Caltech for NASA].

This research makes use of data products from the
Wide-field Infrared Survey Explorer, which is a joint project of the University of California, Los Angeles, and the Jet Propulsion Laboratory/California Institute of Technology, funded by the National Aeronautics and Space Administration. This research also makes use of data products from NEOWISE, which is a project of the Jet Propulsion Laboratory/California Institute of Technology, funded by the Planetary Science Division of the National Aeronautics and Space Administration. This research has made use of the NASA/IPAC Infrared Science Archive, which is operated by the Jet Propulsion Laboratory, California Institute of Technology, under contract with the National Aeronautics and Space Administration. This publication makes use of data products from the Spectro Photometer for the History of the Universe, Epoch of Reionization and Ices Explorer (SPHEREx), which is a joint project of the Jet Propulsion Laboratory and the California Institute of Technology, and is funded by the National Aeronautics and Space Administration.

\end{acknowledgments}

\facilities{WISE, NEOWISE, Spitzer, Palomar/Hale, UKIRT, SPHEREx, 2MASS}

\software{SMDET \citep{SMDET}, WiseView \citep{Caselden_2018}, unTimely Catalog Explorer \citep{ucx}, \texttt{crowdsource} \citep{crowdsource_ascl}, \texttt{unwise\_psf} \citep{unwise_psf}}

\appendix

\section{SPHEREx Spectrophotometry} \label{sec:appendix}

\begin{figure}[!ht]
    \centering
    \includegraphics[width=0.95\textwidth]{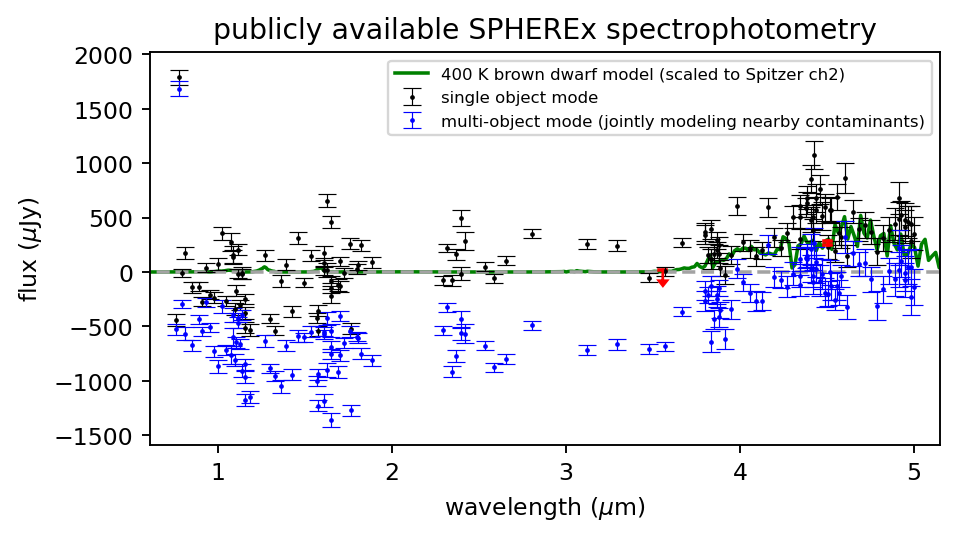}
    \vskip -1em
    \caption{Early SPHEREx spectrophotometry at \objname's per sky pass locations, from the IRSA/SPHEREx Spectrophotometry Tool (black, blue). Spitzer photometry from Deep GLIMPSE is overplotted in red. A $T_{\rm eff} =$ 400~K brown dwarf model spectrum ([m/H]  = 0~dex, log~$g$ = 5.0, C/O = 0.55, log~$K_{zz}$ = 2.0) from the ``LOWZ" model grid, scaled to match the Spitzer ch2 flux, is shown in green \citep{Meisner_2021,Line_LOWZ}.}\label{fig:spherex_spectrum}
\end{figure}

We retrieved SPHEREx spectrophotometry extracted at the relevant \objname~sky locations using the IRSA SPHEREx Spectrophotometry Tool. This SPHEREx spectrophotometry is shown in Figure \ref{fig:spherex_spectrum}. We gathered the SPHEREx spectrophotometry in two ways, referred to as `single object' and `multi-object' mode within the Spectrophotometry Tool's interface. In single object mode, we supplied only \objname's predicted position at the mean epoch of the relevant SPHEREx sky pass ($\S$\ref{sec:spherex}), such that the Spectrophotometry Tool's forced photometry did not attempt to model nearby neighboring sources (we did this separately for each of the two publicly available SPHEREx sky passes). In multi-object mode, we additionally supplied the locations of the three nearest contaminants potentially bright and close enough to meaningfully influence spectrophotometry at \objname's location.

Spectrophotometry at \objname's predicted positions extracted using these two modes differs considerably, as shown in Figure \ref{fig:spherex_spectrum}. Single object mode spectrophotometry is scattered about zero for $\lambda < 2.5$~$\mu$m, albeit with formal uncertainties that for many individual data points imply high statistical significance of deviation from zero flux. At $\lambda > 2.5$~$\mu$m, single object mode spectrophotometry is generally positive though with fluxes too high to match the spectrum of a 400~K brown dwarf scaled to replicate \objname's ch2 brightness. In multi-object mode, the large majority of fluxes extracted at \objname's positions are negative, often with high nominal statistical significance.

Background level estimation in this very crowded Galactic plane field may be a source of systematic uncertainty affecting our extracted SPHEREx spectrophotometry. The Figure \ref{fig:spherex_spectrum} spectrophotometry was extracted using IRSA's default 15 pixel background estimation region. There are two other background estimation region size options available: 21 pixels  and 27 pixels. We re-generated the spectrophotometry for both 21 pixel and 27 pixel background estimation regions, but found negligible overall differences relative to the default. In particular, these two alternative background estimation region sizes did not yield improvements regarding the highly statistically significant negative fluxes in multi-object mode. After taking into account these attempted spectrophotometric measurements, we still conclude that early SPHEREx data do not decisively confirm or rule out the presence of a SPHEREx counterpart to \objname.

\bibliography{sample701}{}

@ARTICLE{Bardalez_2020,
       author = {{Bardalez Gagliuffi}, Daniella C. and {Faherty}, Jacqueline K. and
         {Schneider}, Adam C. and {Meisner}, Aaron and {Caselden}, Dan and
         {Colin}, Guillaume and {Goodman}, Sam and {Kirkpatrick}, J. Davy and
         {Kuchner}, Marc and {Gagn{\'e}}, Jonathan and {Logsdon}, Sarah E. and
         {Burgasser}, Adam J. and {Allers}, Katelyn and {Debes}, John and
         {Wisniewski}, John and {Rothermich}, Austin and
         {Andersen}, Nikolaj Stevnbak and {Th{\'e}venot}, Melina and
         {Walla}, Jim and {Backyard Worlds: Planet 9 Collaboration}},
        title = "{WISEA J083011.95+283716.0: A Missing Link Planetary-mass Object}",
      journal = {ApJ},
         year = 2020,
        month = jun,
       volume = {895},
       number = {2},
          eid = {145},
        pages = {145},
          doi = {10.3847/1538-4357/ab8d25},
archivePrefix = {arXiv},
       eprint = {2004.12829},
 primaryClass = {astro-ph.SR},
       adsurl = {https://ui.adsabs.harvard.edu/abs/2020ApJ...895..145B}
}

@ARTICLE{Brooks_2022,
       author = {{Brooks}, Hunter and {Kirkpatrick}, J. Davy and {Caselden}, Dan and {Schneider}, Adam C. and {Meisner}, Aaron M. and {Faherty}, Jacqueline K. and {Casewell}, S.~L. and {Kuchner}, Marc J. and {Kuchner}, Marc J. and {Backyard Worlds: Planet 9 Collaboration}},
        title = "{Discovery of CWISE J052306.42-015355.4, an Extreme T Subdwarf Candidate}",
      journal = {AJ},
         year = 2022,
        month = feb,
       volume = {163},
       number = {2},
          eid = {47},
        pages = {47},
          doi = {10.3847/1538-3881/ac3a0a},
archivePrefix = {arXiv},
       eprint = {2111.08182},
 primaryClass = {astro-ph.SR},
       adsurl = {https://ui.adsabs.harvard.edu/abs/2022AJ....163...47B}
}

@MISC{Caselden_2018,
       author = {{Caselden}, Dan and {Westin}, Paul, III and {Meisner}, Aaron and {Kuchner}, Marc and {Colin}, Guillaume},
        title = "{WiseView: Visualizing motion and variability of faint WISE sources}",
 howpublished = {Astrophysics Source Code Library, record ascl:1806.004},
         year = 2018,
        month = jun,
          eid = {ascl:1806.004},
        pages = {ascl:1806.004},
archivePrefix = {ascl},
       eprint = {1806.004},
       adsurl = {https://ui.adsabs.harvard.edu/abs/2018ascl.soft06004C}
}

@ARTICLE{Cushing_2011,
       author = {{Cushing}, Michael C. and {Kirkpatrick}, J. Davy and
         {Gelino}, Christopher R. and {Griffith}, Roger L. and
         {Skrutskie}, Michael F. and {Mainzer}, A. and {Marsh}, Kenneth A. and
         {Beichman}, Charles A. and {Burgasser}, Adam J. and {Prato}, Lisa A. and
         {Simcoe}, Robert A. and {Marley}, Mark S. and {Saumon}, D. and
         {Freedman}, Richard S. and {Eisenhardt}, Peter R. and
         {Wright}, Edward L.},
        title = "{The Discovery of Y Dwarfs using Data from the Wide-field Infrared Survey Explorer (WISE)}",
      journal = {ApJ},
         year = 2011,
        month = dec,
       volume = {743},
       number = {1},
          eid = {50},
        pages = {50},
          doi = {10.1088/0004-637X/743/1/50},
archivePrefix = {arXiv},
       eprint = {1108.4678},
 primaryClass = {astro-ph.SR},
       adsurl = {https://ui.adsabs.harvard.edu/abs/2011ApJ...743...50C}
}

@MISC{Faherty_2021,
       author = {{Faherty}, Jacqueline Kelly and {Bardalez Gagliuffi}, Daniella Carolina and {Beichman}, Charles A. and {Burningham}, Ben and {Caselden}, Dan and {Eisenhardt}, Peter and {Gagne}, Jonathan and {Gelino}, Christopher R. and {Gonzales}, Eileen and {Kirkpatrick}, J. Davy and {Kuchner}, Marc Jason and {Marocco}, Federico and {Meisner}, Aaron and {Morley}, Caroline and {Rothermich}, Austin James and {Schneider}, Adam and {Vos}, Johanna and {Whiteford}, Niall},
        title = "{Explaining the Diversity of Cold Worlds}",
 howpublished = {JWST Proposal. Cycle 1, ID. \#2124},
         year = 2021,
        month = mar,
        pages = {2124},
       adsurl = {https://ui.adsabs.harvard.edu/abs/2021jwst.prop.2124F}
}

@ARTICLE{Fazio_2004,
       author = {{Fazio}, G.~G. and {Hora}, J.~L. and {Allen}, L.~E. and
         {Ashby}, M.~L.~N. and {Barmby}, P. and {Deutsch}, L.~K. and
         {Huang}, J. -S. and {Kleiner}, S. and {Marengo}, M. and
         {Megeath}, S.~T. and {Melnick}, G.~J. and {Pahre}, M.~A. and
         {Patten}, B.~M. and {Polizotti}, J. and {Smith}, H.~A. and
         {Taylor}, R.~S. and {Wang}, Z. and {Willner}, S.~P. and
         {Hoffmann}, W.~F. and {Pipher}, J.~L. and {Forrest}, W.~J. and
         {McMurty}, C.~W. and {McCreight}, C.~R. and {McKelvey}, M.~E. and
         {McMurray}, R.~E. and {Koch}, D.~G. and {Moseley}, S.~H. and
         {Arendt}, R.~G. and {Mentzell}, J.~E. and {Marx}, C.~T. and
         {Losch}, P. and {Mayman}, P. and {Eichhorn}, W. and {Krebs}, D. and
         {Jhabvala}, M. and {Gezari}, D.~Y. and {Fixsen}, D.~J. and
         {Flores}, J. and {Shakoorzadeh}, K. and {Jungo}, R. and {Hakun}, C. and
         {Workman}, L. and {Karpati}, G. and {Kichak}, R. and {Whitley}, R. and
         {Mann}, S. and {Tollestrup}, E.~V. and {Eisenhardt}, P. and
         {Stern}, D. and {Gorjian}, V. and {Bhattacharya}, B. and {Carey}, S. and
         {Nelson}, B.~O. and {Glaccum}, W.~J. and {Lacy}, M. and
         {Lowrance}, P.~J. and {Laine}, S. and {Reach}, W.~T. and
         {Stauffer}, J.~A. and {Surace}, J.~A. and {Wilson}, G. and
         {Wright}, E.~L. and {Hoffman}, A. and {Domingo}, G. and {Cohen}, M.},
        title = "{The Infrared Array Camera (IRAC) for the Spitzer Space Telescope}",
      journal = {ApJS},
         year = 2004,
        month = sep,
       volume = {154},
       number = {1},
        pages = {10-17},
          doi = {10.1086/422843},
archivePrefix = {arXiv},
       eprint = {astro-ph/0405616},
 primaryClass = {astro-ph},
       adsurl = {https://ui.adsabs.harvard.edu/abs/2004ApJS..154...10F}
}

@INPROCEEDINGS{gaia_mission,
       author = {{Lindegren}, L. and {Babusiaux}, C. and {Bailer-Jones}, C. and {Bastian}, U. and {Brown}, A.~G.~A. and {Cropper}, M. and {H{\o}g}, E. and {Jordi}, C. and {Katz}, D. and {van Leeuwen}, F. and {Luri}, X. and {Mignard}, F. and {de Bruijne}, J.~H.~J. and {Prusti}, T.},
        title = "{The Gaia mission: science, organization and present status}",
    booktitle = {A Giant Step: from Milli- to Micro-arcsecond Astrometry},
         year = 2008,
       editor = {{Jin}, W.~J. and {Platais}, I. and {Perryman}, M.~A.~C.},
       volume = {248},
        month = jul,
        pages = {217-223},
          doi = {10.1017/S1743921308019133},
       adsurl = {https://ui.adsabs.harvard.edu/abs/2008IAUS..248..217L}
}

@ARTICLE{Jones_1972,
       author = {{Jones}, Eric M.},
        title = "{Reduced-Proper Diagrams}",
      journal = {ApJ},
         year = 1972,
        month = may,
       volume = {173},
        pages = {671},
          doi = {10.1086/151454},
       adsurl = {https://ui.adsabs.harvard.edu/abs/1972ApJ...173..671J}
}

@ARTICLE{Kirkpatrick_2011,
       author = {{Kirkpatrick}, J. Davy and {Cushing}, Michael C. and
         {Gelino}, Christopher R. and {Griffith}, Roger L. and
         {Skrutskie}, Michael F. and {Marsh}, Kenneth A. and
         {Wright}, Edward L. and {Mainzer}, A. and {Eisenhardt}, Peter R. and
         {McLean}, Ian S. and {Thompson}, Maggie A. and {Bauer}, James M. and
         {Benford}, Dominic J. and {Bridge}, Carrie R. and {Lake}, Sean E. and
         {Petty}, Sara M. and {Stanford}, S.~A. and {Tsai}, Chao-Wei and
         {Bailey}, Vanessa and {Beichman}, Charles A. and {Bloom}, Joshua S. and
         {Bochanski}, John J. and {Burgasser}, Adam J. and {Capak}, Peter L. and
         {Cruz}, Kelle L. and {Hinz}, Philip M. and {Kartaltepe}, Jeyhan S. and
         {Knox}, Russell P. and {Manohar}, Swarnima and {Masters}, Daniel and
         {Morales-Calder{\'o}n}, Maria and {Prato}, Lisa A. and
         {Rodigas}, Timothy J. and {Salvato}, Mara and {Schurr}, Steven D. and
         {Scoville}, Nicholas Z. and {Simcoe}, Robert A. and
         {Stapelfeldt}, Karl R. and {Stern}, Daniel and {Stock}, Nathan D. and
         {Vacca}, William D.},
        title = "{The First Hundred Brown Dwarfs Discovered by the Wide-field Infrared Survey Explorer (WISE)}",
      journal = {ApJS},
         year = 2011,
        month = dec,
       volume = {197},
       number = {2},
          eid = {19},
        pages = {19},
          doi = {10.1088/0067-0049/197/2/19},
archivePrefix = {arXiv},
       eprint = {1108.4677},
 primaryClass = {astro-ph.SR},
       adsurl = {https://ui.adsabs.harvard.edu/abs/2011ApJS..197...19K}
}

@ARTICLE{Kirkpatrick_2012,
       author = {{Kirkpatrick}, J. Davy and {Gelino}, Christopher R. and
         {Cushing}, Michael C. and {Mace}, Gregory N. and {Griffith}, Roger L. and
         {Skrutskie}, Michael F. and {Marsh}, Kenneth A. and
         {Wright}, Edward L. and {Eisenhardt}, Peter R. and {McLean}, Ian S. and
         {Mainzer}, Amanda K. and {Burgasser}, Adam J. and {Tinney}, C.~G. and
         {Parker}, Stephen and {Salter}, Graeme},
        title = "{Further Defining Spectral Type ``Y'' and Exploring the Low-mass End of the Field Brown Dwarf Mass Function}",
      journal = {ApJ},
         year = 2012,
        month = jul,
       volume = {753},
       number = {2},
          eid = {156},
        pages = {156},
          doi = {10.1088/0004-637X/753/2/156},
archivePrefix = {arXiv},
       eprint = {1205.2122},
 primaryClass = {astro-ph.SR},
       adsurl = {https://ui.adsabs.harvard.edu/abs/2012ApJ...753..156K}
}

@ARTICLE{Kirkpatrick_2021a,
       author = {{Kirkpatrick}, J. Davy and {Gelino}, Christopher R. and {Faherty}, Jacqueline K. and {Meisner}, Aaron M. and {Caselden}, Dan and {Schneider}, Adam C. and {Marocco}, Federico and {Cayago}, Alfred J. and {Smart}, R.~L. and {Eisenhardt}, Peter R. and {Kuchner}, Marc J. and {Wright}, Edward L. and {Cushing}, Michael C. and {Allers}, Katelyn N. and {Bardalez Gagliuffi}, Daniella C. and {Burgasser}, Adam J. and {Gagn{\'e}}, Jonathan and {Logsdon}, Sarah E. and {Martin}, Emily C. and {Ingalls}, James G. and {Lowrance}, Patrick J. and {Abrahams}, Ellianna S. and {Aganze}, Christian and {Gerasimov}, Roman and {Gonzales}, Eileen C. and {Hsu}, Chih-Chun and {Kamraj}, Nikita and {Kiman}, Rocio and {Rees}, Jon and {Theissen}, Christopher and {Ammar}, Kareem and {Andersen}, Nikolaj Stevnbak and {Beaulieu}, Paul and {Colin}, Guillaume and {Elachi}, Charles A. and {Goodman}, Samuel J. and {Gramaize}, L{\'e}opold and {Hamlet}, Leslie K. and {Hong}, Justin and {Jonkeren}, Alexander and {Khalil}, Mohammed and {Martin}, David W. and {Pendrill}, William and {Pumphrey}, Benjamin and {Rothermich}, Austin and {Sainio}, Arttu and {Stenner}, Andres and {Tanner}, Christopher and {Th{\'e}venot}, Melina and {Voloshin}, Nikita V. and {Walla}, Jim and {W{\k{e}}dracki}, Zbigniew and {Backyard Worlds: Planet 9 Collaboration}},
        title = "{The Field Substellar Mass Function Based on the Full-sky 20 pc Census of 525 L, T, and Y Dwarfs}",
      journal = {ApJS},
         year = 2021,
        month = mar,
       volume = {253},
       number = {1},
          eid = {7},
        pages = {7},
          doi = {10.3847/1538-4365/abd107},
archivePrefix = {arXiv},
       eprint = {2011.11616},
 primaryClass = {astro-ph.SR},
       adsurl = {https://ui.adsabs.harvard.edu/abs/2021ApJS..253....7K}
}

@ARTICLE{Leggett_2017,
       author = {{Leggett}, S.~K. and {Tremblin}, P. and {Esplin}, T.~L. and
         {Luhman}, K.~L. and {Morley}, Caroline V.},
        title = "{The Y-type Brown Dwarfs: Estimates of Mass and Age from New Astrometry, Homogenized Photometry, and Near-infrared Spectroscopy}",
      journal = {ApJ},
         year = 2017,
        month = jun,
       volume = {842},
       number = {2},
          eid = {118},
        pages = {118},
          doi = {10.3847/1538-4357/aa6fb5},
archivePrefix = {arXiv},
       eprint = {1704.03573},
 primaryClass = {astro-ph.SR},
       adsurl = {https://ui.adsabs.harvard.edu/abs/2017ApJ...842..118L}
}

@ARTICLE{Leggett_2019b,
       author = {{Leggett}, Sandy and {Apai}, Daniel and {Burgasser}, Adam and {Cushing}, Michael and {Dupuy}, Trent and {Faherty}, Jackie and {Gizis}, John and {Kirkpatrick}, J. Davy and {Marley}, Mark and {Morley}, Caroline and {Schneider}, Adam and {Sousa-Silva}, Clara},
        title = "{Discovery of Cold Brown Dwarfs or Free-Floating Giant Planets Close to the Sun}",
      journal = {BAAS},
         year = 2019,
        month = may,
       volume = {51},
       number = {3},
          eid = {95},
        pages = {95},
          doi = {10.48550/arXiv.1903.04686},
archivePrefix = {arXiv},
       eprint = {1903.04686},
 primaryClass = {astro-ph.SR},
       adsurl = {https://ui.adsabs.harvard.edu/abs/2019BAAS...51c..95L}
}

@ARTICLE{Luhman_2014,
       author = {{Luhman}, K.~L.},
        title = "{Discovery of a \raisebox{-0.5ex}\textasciitilde250 K Brown Dwarf at 2 pc from the Sun}",
      journal = {ApJL},
         year = 2014,
        month = may,
       volume = {786},
       number = {2},
          eid = {L18},
        pages = {L18},
          doi = {10.1088/2041-8205/786/2/L18},
archivePrefix = {arXiv},
       eprint = {1404.6501},
 primaryClass = {astro-ph.GA},
       adsurl = {https://ui.adsabs.harvard.edu/abs/2014ApJ...786L..18L}
}

@ARTICLE{Mainzer_2011,
       author = {{Mainzer}, A. and {Cushing}, Michael C. and {Skrutskie}, M. and
         {Gelino}, C.~R. and {Kirkpatrick}, J. Davy and {Jarrett}, T. and
         {Masci}, F. and {Marley}, Mark S. and {Saumon}, D. and {Wright}, E. and
         {Beaton}, R. and {Dietrich}, M. and {Eisenhardt}, P. and
         {Garnavich}, P. and {Kuhn}, O. and {Leisawitz}, D. and {Marsh}, K. and
         {McLean}, I. and {Padgett}, D. and {Rueff}, K.},
        title = "{The First Ultra-cool Brown Dwarf Discovered by the Wide-field Infrared Survey Explorer}",
      journal = {ApJ},
         year = 2011,
        month = jan,
       volume = {726},
       number = {1},
          eid = {30},
        pages = {30},
          doi = {10.1088/0004-637X/726/1/30},
archivePrefix = {arXiv},
       eprint = {1011.2279},
 primaryClass = {astro-ph.GA},
       adsurl = {https://ui.adsabs.harvard.edu/abs/2011ApJ...726...30M}
}

@ARTICLE{Marocco_2019,
       author = {{Marocco}, Federico and {Caselden}, Dan and {Meisner}, Aaron M. and
         {Kirkpatrick}, J. Davy and {Wright}, Edward L. and
         {Faherty}, Jacqueline K. and {Gelino}, Christopher R. and
         {Eisenhardt}, Peter R.~M. and {Fowler}, John W. and
         {Cushing}, Michael C. and {Cutri}, Roc M. and {Garcia}, Nelson and
         {Jarrett}, Thomas H. and {Koontz}, Renata and {Mainzer}, Amanda and
         {Marchese}, Elijah J. and {Mobasher}, Bahram and {Schlegel}, David J. and
         {Stern}, Daniel and {Teplitz}, Harry I.},
        title = "{CWISEP J193518.59-154620.3: An Extremely Cold Brown Dwarf in the Solar Neighborhood Discovered with CatWISE}",
      journal = {ApJ},
         year = 2019,
        month = aug,
       volume = {881},
       number = {1},
          eid = {17},
        pages = {17},
          doi = {10.3847/1538-4357/ab2bf0},
archivePrefix = {arXiv},
       eprint = {1906.08913},
 primaryClass = {astro-ph.SR},
       adsurl = {https://ui.adsabs.harvard.edu/abs/2019ApJ...881...17M}
}

@ARTICLE{Meisner_2020b,
       author = {{Meisner}, Aaron M. and {Faherty}, Jacqueline K. and {Kirkpatrick}, J. Davy and {Schneider}, Adam C. and {Caselden}, Dan and {Gagn{\'e}}, Jonathan and {Kuchner}, Marc J. and {Burgasser}, Adam J. and {Casewell}, Sarah L. and {Debes}, John H. and {Artigau}, {\'E}tienne and {Bardalez Gagliuffi}, Daniella C. and {Logsdon}, Sarah E. and {Kiman}, Rocio and {Allers}, Katelyn and {Hsu}, Chih-chun and {Wisniewski}, John P. and {Allen}, Michaela B. and {Beaulieu}, Paul and {Colin}, Guillaume and {Durantini Luca}, Hugo A. and {Goodman}, Sam and {Gramaize}, L{\'e}opold and {Hamlet}, Leslie K. and {Hinckley}, Ken and {Kiwy}, Frank and {Martin}, David W. and {Pendrill}, William and {Rothermich}, Austin and {Sainio}, Arttu and {Sch{\"u}mann}, J{\"o}rg and {Andersen}, Nikolaj Stevnbak and {Tanner}, Christopher and {Thakur}, Vinod and {Th{\'e}venot}, Melina and {Walla}, Jim and {W{\k{e}}dracki}, Zbigniew and {Aganze}, Christian and {Gerasimov}, Roman and {Theissen}, Christopher and {Backyard Worlds: Planet 9 Collaboration}},
        title = "{Spitzer Follow-up of Extremely Cold Brown Dwarfs Discovered by the Backyard Worlds: Planet 9 Citizen Science Project}",
      journal = {ApJ},
         year = 2020,
        month = aug,
       volume = {899},
       number = {2},
          eid = {123},
        pages = {123},
          doi = {10.3847/1538-4357/aba633},
archivePrefix = {arXiv},
       eprint = {2008.06396},
 primaryClass = {astro-ph.SR},
       adsurl = {https://ui.adsabs.harvard.edu/abs/2020ApJ...899..123M}
}

@ARTICLE{Meisner_2021,
       author = {{Meisner}, Aaron M. and {Schneider}, Adam C. and {Burgasser}, Adam J. and {Marocco}, Federico and {Line}, Michael R. and {Faherty}, Jacqueline K. and {Kirkpatrick}, J. Davy and {Caselden}, Dan and {Kuchner}, Marc J. and {Gelino}, Christopher R. and {Gagn{\'e}}, Jonathan and {Theissen}, Christopher and {Gerasimov}, Roman and {Aganze}, Christian and {Hsu}, Chih-chun and {Wisniewski}, John P. and {Casewell}, Sarah L. and {Bardalez Gagliuffi}, Daniella C. and {Logsdon}, Sarah E. and {Eisenhardt}, Peter R.~M. and {Allers}, Katelyn and {Debes}, John H. and {Allen}, Michaela B. and {Stevnbak Andersen}, Nikolaj and {Goodman}, Sam and {Gramaize}, L{\'e}opold and {Martin}, David W. and {Sainio}, Arttu and {Cushing}, Michael C. and {Backyard Worlds: Planet 9 Collaboration}},
        title = "{New Candidate Extreme T Subdwarfs from the Backyard Worlds: Planet 9 Citizen Science Project}",
      journal = {ApJ},
         year = 2021,
        month = jul,
       volume = {915},
       number = {2},
          eid = {120},
        pages = {120},
          doi = {10.3847/1538-4357/ac013c},
archivePrefix = {arXiv},
       eprint = {2106.01387},
 primaryClass = {astro-ph.SR},
       adsurl = {https://ui.adsabs.harvard.edu/abs/2021ApJ...915..120M}
}

@ARTICLE{Meisner_2022,
       author = {{Meisner}, A.~M. and {Caselden}, D. and {Schlafly}, E.~F. and {Kiwy}, F.},
        title = "{unTimely: a Full-sky, Time-Domain unWISE Catalog}",
      journal = {arXiv e-prints},
         year = 2022,
        month = sep,
          eid = {arXiv:2209.14327},
        pages = {arXiv:2209.14327},
archivePrefix = {arXiv},
       eprint = {2209.14327},
 primaryClass = {astro-ph.IM},
       adsurl = {https://ui.adsabs.harvard.edu/abs/2022arXiv220914327M}
}

@ARTICLE{Meisner_2023,
       author = {{Meisner}, Aaron M. and {Leggett}, S.~K. and {Logsdon}, Sarah E. and {Schneider}, Adam C. and {Tremblin}, Pascal and {Phillips}, Mark},
        title = "{Exploring the Extremes: Characterizing a New Population of Old and Cold Brown Dwarfs}",
      journal = {AJ},
         year = 2023,
        month = aug,
       volume = {166},
       number = {2},
          eid = {57},
        pages = {57},
          doi = {10.3847/1538-3881/acdb68},
archivePrefix = {arXiv},
       eprint = {2301.09817},
 primaryClass = {astro-ph.SR},
       adsurl = {https://ui.adsabs.harvard.edu/abs/2023AJ....166...57M}
}

@ARTICLE{Morley_2012,
       author = {{Morley}, Caroline V. and {Fortney}, Jonathan J. and {Marley}, Mark S. and
         {Visscher}, Channon and {Saumon}, Didier and {Leggett}, S.~K.},
        title = "{Neglected Clouds in T and Y Dwarf Atmospheres}",
      journal = {ApJ},
         year = 2012,
        month = sep,
       volume = {756},
       number = {2},
          eid = {172},
        pages = {172},
          doi = {10.1088/0004-637X/756/2/172},
archivePrefix = {arXiv},
       eprint = {1206.4313},
 primaryClass = {astro-ph.SR},
       adsurl = {https://ui.adsabs.harvard.edu/abs/2012ApJ...756..172M}
}

@ARTICLE{Morley_2014,
       author = {{Morley}, Caroline V. and {Marley}, Mark S. and {Fortney}, Jonathan J. and
         {Lupu}, Roxana and {Saumon}, Didier and {Greene}, Tom and
         {Lodders}, Katharina},
        title = "{Water Clouds in Y Dwarfs and Exoplanets}",
      journal = {ApJ},
         year = 2014,
        month = may,
       volume = {787},
       number = {1},
          eid = {78},
        pages = {78},
          doi = {10.1088/0004-637X/787/1/78},
archivePrefix = {arXiv},
       eprint = {1404.0005},
 primaryClass = {astro-ph.SR},
       adsurl = {https://ui.adsabs.harvard.edu/abs/2014ApJ...787...78M}
}

@ARTICLE{Robbins_2023,
       author = {{Robbins}, Grady and {Meisner}, Aaron M. and {Schneider}, Adam C. and {Burgasser}, Adam J. and {Kirkpatrick}, J. Davy and {Gagn{\'e}}, Jonathan and {Hsu}, Chih-Chun and {Moranta}, Leslie and {Casewell}, Sarah and {Marocco}, Federico and {Gerasimov}, Roman and {Faherty}, Jacqueline K. and {Kuchner}, Marc J. and {Caselden}, Dan and {Cushing}, Michael C. and {Alejandro}, Sherelyn and {Backyard Worlds: Cool Neighbors Collaboration}},
        title = "{CWISE J105512.11+544328.3: A Nearby Y Dwarf Spectroscopically Confirmed with Keck/NIRES}",
      journal = {ApJ},
         year = 2023,
        month = nov,
       volume = {958},
       number = {1},
          eid = {94},
        pages = {94},
          doi = {10.3847/1538-4357/ad0043},
archivePrefix = {arXiv},
       eprint = {2310.09524},
 primaryClass = {astro-ph.SR},
       adsurl = {https://ui.adsabs.harvard.edu/abs/2023ApJ...958...94R}
}

@ARTICLE{Schneider_2020,
       author = {{Schneider}, Adam C. and {Burgasser}, Adam J. and {Gerasimov}, Roman and {Marocco}, Federico and {Gagn{\'e}}, Jonathan and {Goodman}, Sam and {Beaulieu}, Paul and {Pendrill}, William and {Rothermich}, Austin and {Sainio}, Arttu and {Kuchner}, Marc J. and {Caselden}, Dan and {Meisner}, Aaron M. and {Faherty}, Jacqueline K. and {Mamajek}, Eric E. and {Hsu}, Chih-Chun and {Greco}, Jennifer J. and {Cushing}, Michael C. and {Kirkpatrick}, J. Davy and {Bardalez-Gagliuffi}, Daniella and {Logsdon}, Sarah E. and {Allers}, Katelyn and {Debes}, John H. and {Backyard Worlds: Planet 9 Collaboration}},
        title = "{WISEA J041451.67-585456.7 and WISEA J181006.18-101000.5: The First Extreme T-type Subdwarfs?}",
      journal = {ApJ},
         year = 2020,
        month = jul,
       volume = {898},
       number = {1},
          eid = {77},
        pages = {77},
          doi = {10.3847/1538-4357/ab9a40},
archivePrefix = {arXiv},
       eprint = {2007.03836},
 primaryClass = {astro-ph.SR},
       adsurl = {https://ui.adsabs.harvard.edu/abs/2020ApJ...898...77S}
}

@ARTICLE{UKIDSS,
       author = {{Lawrence}, A. and {Warren}, S.~J. and {Almaini}, O. and {Edge}, A.~C. and
         {Hambly}, N.~C. and {Jameson}, R.~F. and {Lucas}, P. and {Casali}, M. and
         {Adamson}, A. and {Dye}, S. and {Emerson}, J.~P. and {Foucaud}, S. and
         {Hewett}, P. and {Hirst}, P. and {Hodgkin}, S.~T. and {Irwin}, M.~J. and
         {Lodieu}, N. and {McMahon}, R.~G. and {Simpson}, C. and {Smail}, I. and
         {Mortlock}, D. and {Folger}, M.},
        title = "{The UKIRT Infrared Deep Sky Survey (UKIDSS)}",
      journal = {MNRAS},
         year = 2007,
        month = aug,
       volume = {379},
       number = {4},
        pages = {1599-1617},
          doi = {10.1111/j.1365-2966.2007.12040.x},
archivePrefix = {arXiv},
       eprint = {astro-ph/0604426},
 primaryClass = {astro-ph},
       adsurl = {https://ui.adsabs.harvard.edu/abs/2007MNRAS.379.1599L}
}

@ARTICLE{Werner_2004,
       author = {{Werner}, M.~W. and {Roellig}, T.~L. and {Low}, F.~J. and
         {Rieke}, G.~H. and {Rieke}, M. and {Hoffmann}, W.~F. and {Young}, E. and
         {Houck}, J.~R. and {Brandl}, B. and {Fazio}, G.~G. and {Hora}, J.~L. and
         {Gehrz}, R.~D. and {Helou}, G. and {Soifer}, B.~T. and {Stauffer}, J. and
         {Keene}, J. and {Eisenhardt}, P. and {Gallagher}, D. and
         {Gautier}, T.~N. and {Irace}, W. and {Lawrence}, C.~R. and
         {Simmons}, L. and {Van Cleve}, J.~E. and {Jura}, M. and
         {Wright}, E.~L. and {Cruikshank}, D.~P.},
        title = "{The Spitzer Space Telescope Mission}",
      journal = {ApJS},
         year = 2004,
        month = sep,
       volume = {154},
       number = {1},
        pages = {1-9},
          doi = {10.1086/422992},
archivePrefix = {arXiv},
       eprint = {astro-ph/0406223},
 primaryClass = {astro-ph},
       adsurl = {https://ui.adsabs.harvard.edu/abs/2004ApJS..154....1W}
}

@ARTICLE{Wright_2010,
       author = {{Wright}, Edward L. and {Eisenhardt}, Peter R.~M. and {Mainzer}, Amy K. and
         {Ressler}, Michael E. and {Cutri}, Roc M. and {Jarrett}, Thomas and
         {Kirkpatrick}, J. Davy and {Padgett}, Deborah and
         {McMillan}, Robert S. and {Skrutskie}, Michael and {Stanford}, S.~A. and
         {Cohen}, Martin and {Walker}, Russell G. and {Mather}, John C. and
         {Leisawitz}, David and {Gautier}, Thomas N., III and {McLean}, Ian and
         {Benford}, Dominic and {Lonsdale}, Carol J. and {Blain}, Andrew and
         {Mendez}, Bryan and {Irace}, William R. and {Duval}, Valerie and
         {Liu}, Fengchuan and {Royer}, Don and {Heinrichsen}, Ingolf and
         {Howard}, Joan and {Shannon}, Mark and {Kendall}, Martha and
         {Walsh}, Amy L. and {Larsen}, Mark and {Cardon}, Joel G. and
         {Schick}, Scott and {Schwalm}, Mark and {Abid}, Mohamed and
         {Fabinsky}, Beth and {Naes}, Larry and {Tsai}, Chao-Wei},
        title = "{The Wide-field Infrared Survey Explorer (WISE): Mission Description and Initial On-orbit Performance}",
      journal = {AJ},
         year = 2010,
        month = dec,
       volume = {140},
       number = {6},
        pages = {1868-1881},
          doi = {10.1088/0004-6256/140/6/1868},
archivePrefix = {arXiv},
       eprint = {1008.0031},
 primaryClass = {astro-ph.IM},
       adsurl = {https://ui.adsabs.harvard.edu/abs/2010AJ....140.1868W}
}

@ARTICLE{Kirkpatrick_2024,
       author = {{Kirkpatrick}, J. Davy and {Marocco}, Federico and {Gelino}, Christopher R. and {Raghu}, Yadukrishna and {Faherty}, Jacqueline K. and {Bardalez Gagliuffi}, Daniella C. and {Schurr}, Steven D. and {Apps}, Kevin and {Schneider}, Adam C. and {Meisner}, Aaron M. and {Kuchner}, Marc J. and {Caselden}, Dan and {Smart}, R.~L. and {Casewell}, S.~L. and {Raddi}, Roberto and {Kesseli}, Aurora and {Stevnbak Andersen}, Nikolaj and {Antonini}, Edoardo and {Beaulieu}, Paul and {Bickle}, Thomas P. and {Bilsing}, Martin and {Chieng}, Raymond and {Colin}, Guillaume and {Deen}, Sam and {Dereveanco}, Alexandru and {Doll}, Katharina and {Durantini Luca}, Hugo A. and {Frazer}, Anya and {Gantier}, Jean Marc and {Gramaize}, L{\'e}opold and {Grant}, Kristin and {Hamlet}, Leslie K. and {Higashimura}, Hiro and {Hyogo}, Michiharu and {Ja{\l}owiczor}, Peter A. and {Jonkeren}, Alexander and {Kabatnik}, Martin and {Kiwy}, Frank and {Martin}, David W. and {Michaels}, Marianne N. and {Pendrill}, William and {Pessanha Machado}, Celso and {Pumphrey}, Benjamin and {Rothermich}, Austin and {Russwurm}, Rebekah and {Sainio}, Arttu and {Sanchez}, John and {Sapelkin-Tambling}, Fyodor Theo and {Sch{\"u}mann}, J{\"o}rg and {Selg-Mann}, Karl and {Singh}, Harshdeep and {Stenner}, Andres and {Sun}, Guoyou and {Tanner}, Christopher and {Th{\'e}venot}, Melina and {Ventura}, Maurizio and {Voloshin}, Nikita V. and {Walla}, Jim and {W{\k{e}}dracki}, Zbigniew and {Adorno}, Jose I. and {Aganze}, Christian and {Allers}, Katelyn N. and {Brooks}, Hunter and {Burgasser}, Adam J. and {Calamari}, Emily and {Connor}, Thomas and {Costa}, Edgardo and {Eisenhardt}, Peter R. and {Gagn{\'e}}, Jonathan and {Gerasimov}, Roman and {Gonzales}, Eileen C. and {Hsu}, Chih-Chun and {Kiman}, Rocio and {Li}, Guodong and {Low}, Ryan and {Mamajek}, Eric and {Pantoja}, Blake M. and {Popinchalk}, Mark and {Rees}, Jon M. and {Stern}, Daniel and {Su{\'a}rez}, Genaro and {Theissen}, Christopher and {Tsai}, Chao-Wei and {Vos}, Johanna M. and {Zurek}, David and {The Backyard Worlds: Planet 9 Collaboration}},
        title = "{The Initial Mass Function Based on the Full-sky 20 pc Census of {\ensuremath{\sim}}3600 Stars and Brown Dwarfs}",
      journal = {ApJS},
         year = 2024,
        month = apr,
       volume = {271},
       number = {2},
          eid = {55},
        pages = {55},
          doi = {10.3847/1538-4365/ad24e2},
archivePrefix = {arXiv},
       eprint = {2312.03639},
 primaryClass = {astro-ph.SR},
       adsurl = {https://ui.adsabs.harvard.edu/abs/2024ApJS..271...55K}
}

@ARTICLE{Calissendorff_2023,
       author = {{Calissendorff}, Per and {De Furio}, Matthew and {Meyer}, Michael and {Albert}, Lo{\"\i}c and {Aganze}, Christian and {Ali-Dib}, Mohamad and {Bardalez Gagliuffi}, Daniella C. and {Baron}, Frederique and {Beichman}, Charles A. and {Burgasser}, Adam J. and {Cushing}, Michael C. and {Faherty}, Jacqueline Kelly and {Fontanive}, Cl{\'e}mence and {Gelino}, Christopher R. and {Gizis}, John E. and {Greenbaum}, Alexandra Z. and {Kirkpatrick}, J. Davy and {Leggett}, Sandy K. and {Martinache}, Frantz and {Mary}, David and {N'Diaye}, Mamadou and {Pope}, Benjamin J.~S. and {Roellig}, Thomas and {Sahlmann}, Johannes and {Sivaramakrishnan}, Anand and {Thorngren}, Daniel Peter and {Ygouf}, Marie and {Vandal}, Thomas},
        title = "{JWST/NIRCam Discovery of the First Y+Y Brown Dwarf Binary: WISE J033605.05-014350.4}",
      journal = {ApJL},
         year = 2023,
        month = apr,
       volume = {947},
       number = {2},
          eid = {L30},
        pages = {L30},
          doi = {10.3847/2041-8213/acc86d},
archivePrefix = {arXiv},
       eprint = {2303.16923},
 primaryClass = {astro-ph.SR},
       adsurl = {https://ui.adsabs.harvard.edu/abs/2023ApJ...947L..30C}
}

@ARTICLE{Beiler_sample,
       author = {{Beiler}, Samuel A. and {Cushing}, Michael C. and {Kirkpatrick}, J. Davy and {Schneider}, Adam C. and {Mukherjee}, Sagnick and {Marley}, Mark S. and {Marocco}, Federico and {Smart}, Richard L.},
        title = "{Precise Bolometric Luminosities and Effective Temperatures of 23 Late-T and Y Dwarfs Obtained with JWST}",
      journal = {ApJ},
         year = 2024,
        month = oct,
       volume = {973},
       number = {2},
          eid = {107},
        pages = {107},
          doi = {10.3847/1538-4357/ad6301},
archivePrefix = {arXiv},
       eprint = {2407.08518},
 primaryClass = {astro-ph.SR},
       adsurl = {https://ui.adsabs.harvard.edu/abs/2024ApJ...973..107B}
}

@INPROCEEDINGS{SMDET,
       author = {{Caselden}, D. and {Colin}, G. and {Lack}, L. and {Marocco}, F. and {Kirkpatrick}, J. and {Meisner}, A.},
        title = "{WISE Image Cubes, Neural Nets, and Moving Objects with SMDET}",
    booktitle = {American Astronomical Society Meeting Abstracts \#235},
         year = 2020,
       series = {American Astronomical Society Meeting Abstracts},
       volume = {235},
        month = jan,
          eid = {274.18},
        pages = {274.18},
       adsurl = {https://ui.adsabs.harvard.edu/abs/2020AAS...23527418C}
}

@ARTICLE{Gaia_DR3,
       author = {{Gaia Collaboration} and {Brown}, A.~G.~A. and {Vallenari}, A. and {Prusti}, T. and {de Bruijne}, J.~H.~J. and {Babusiaux}, C. and {Biermann}, M. and {Creevey}, O.~L. and {Evans}, D.~W. and {Eyer}, L. and {Hutton}, A. and {Jansen}, F. and {Jordi}, C. and {Klioner}, S.~A. and {Lammers}, U. and {Lindegren}, L. and {Luri}, X. and {Mignard}, F. and {Panem}, C. and {Pourbaix}, D. and {Randich}, S. and {Sartoretti}, P. and {Soubiran}, C. and {Walton}, N.~A. and {Arenou}, F. and {Bailer-Jones}, C.~A.~L. and {Bastian}, U. and {Cropper}, M. and {Drimmel}, R. and {Katz}, D. and {Lattanzi}, M.~G. and {van Leeuwen}, F. and {Bakker}, J. and {Cacciari}, C. and {Casta{\~n}eda}, J. and {De Angeli}, F. and {Ducourant}, C. and {Fabricius}, C. and {Fouesneau}, M. and {Fr{\'e}mat}, Y. and {Guerra}, R. and {Guerrier}, A. and {Guiraud}, J. and {Jean-Antoine Piccolo}, A. and {Masana}, E. and {Messineo}, R. and {Mowlavi}, N. and {Nicolas}, C. and {Nienartowicz}, K. and {Pailler}, F. and {Panuzzo}, P. and {Riclet}, F. and {Roux}, W. and {Seabroke}, G.~M. and {Sordo}, R. and {Tanga}, P. and {Th{\'e}venin}, F. and {Gracia-Abril}, G. and {Portell}, J. and {Teyssier}, D. and {Altmann}, M. and {Andrae}, R. and {Bellas-Velidis}, I. and {Benson}, K. and {Berthier}, J. and {Blomme}, R. and {Brugaletta}, E. and {Burgess}, P.~W. and {Busso}, G. and {Carry}, B. and {Cellino}, A. and {Cheek}, N. and {Clementini}, G. and {Damerdji}, Y. and {Davidson}, M. and {Delchambre}, L. and {Dell'Oro}, A. and {Fern{\'a}ndez-Hern{\'a}ndez}, J. and {Galluccio}, L. and {Garc{\'\i}a-Lario}, P. and {Garcia-Reinaldos}, M. and {Gonz{\'a}lez-N{\'u}{\~n}ez}, J. and {Gosset}, E. and {Haigron}, R. and {Halbwachs}, J. -L. and {Hambly}, N.~C. and {Harrison}, D.~L. and {Hatzidimitriou}, D. and {Heiter}, U. and {Hern{\'a}ndez}, J. and {Hestroffer}, D. and {Hodgkin}, S.~T. and {Holl}, B. and {Jan{\ss}en}, K. and {Jevardat de Fombelle}, G. and {Jordan}, S. and {Krone-Martins}, A. and {Lanzafame}, A.~C. and {L{\"o}ffler}, W. and {Lorca}, A. and {Manteiga}, M. and {Marchal}, O. and {Marrese}, P.~M. and {Moitinho}, A. and {Mora}, A. and {Muinonen}, K. and {Osborne}, P. and {Pancino}, E. and {Pauwels}, T. and {Petit}, J. -M. and {Recio-Blanco}, A. and {Richards}, P.~J. and {Riello}, M. and {Rimoldini}, L. and {Robin}, A.~C. and {Roegiers}, T. and {Rybizki}, J. and {Sarro}, L.~M. and {Siopis}, C. and {Smith}, M. and {Sozzetti}, A. and {Ulla}, A. and {Utrilla}, E. and {van Leeuwen}, M. and {van Reeven}, W. and {Abbas}, U. and {Abreu Aramburu}, A. and {Accart}, S. and {Aerts}, C. and {Aguado}, J.~J. and {Ajaj}, M. and {Altavilla}, G. and {{\'A}lvarez}, M.~A. and {{\'A}lvarez Cid-Fuentes}, J. and {Alves}, J. and {Anderson}, R.~I. and {Anglada Varela}, E. and {Antoja}, T. and {Audard}, M. and {Baines}, D. and {Baker}, S.~G. and {Balaguer-N{\'u}{\~n}ez}, L. and {Balbinot}, E. and {Balog}, Z. and {Barache}, C. and {Barbato}, D. and {Barros}, M. and {Barstow}, M.~A. and {Bartolom{\'e}}, S. and {Bassilana}, J. -L. and {Bauchet}, N. and {Baudesson-Stella}, A. and {Becciani}, U. and {Bellazzini}, M. and {Bernet}, M. and {Bertone}, S. and {Bianchi}, L. and {Blanco-Cuaresma}, S. and {Boch}, T. and {Bombrun}, A. and {Bossini}, D. and {Bouquillon}, S. and {Bragaglia}, A. and {Bramante}, L. and {Breedt}, E. and {Bressan}, A. and {Brouillet}, N. and {Bucciarelli}, B. and {Burlacu}, A. and {Busonero}, D. and {Butkevich}, A.~G. and {Buzzi}, R. and {Caffau}, E. and {Cancelliere}, R. and {C{\'a}novas}, H. and {Cantat-Gaudin}, T. and {Carballo}, R. and {Carlucci}, T. and {Carnerero}, M.~I. and {Carrasco}, J.~M. and {Casamiquela}, L. and {Castellani}, M. and {Castro-Ginard}, A. and {Castro Sampol}, P. and {Chaoul}, L. and {Charlot}, P. and {Chemin}, L. and {Chiavassa}, A. and {Cioni}, M. -R.~L. and {Comoretto}, G. and {Cooper}, W.~J. and {Cornez}, T. and {Cowell}, S. and {Crifo}, F. and {Crosta}, M. and {Crowley}, C. and {Dafonte}, C. and {Dapergolas}, A. and {David}, M. and {David}, P. and {de Laverny}, P. and {De Luise}, F. and {De March}, R. and {De Ridder}, J. and {de Souza}, R. and {de Teodoro}, P. and {de Torres}, A. and {del Peloso}, E.~F. and {del Pozo}, E. and {Delbo}, M. and {Delgado}, A. and {Delgado}, H.~E. and {Delisle}, J. -B. and {Di Matteo}, P. and {Diakite}, S. and {Diener}, C. and {Distefano}, E. and {Dolding}, C. and {Eappachen}, D. and {Edvardsson}, B. and {Enke}, H. and {Esquej}, P. and {Fabre}, C. and {Fabrizio}, M. and {Faigler}, S. and {Fedorets}, G. and {Fernique}, P. and {Fienga}, A. and {Figueras}, F. and {Fouron}, C. and {Fragkoudi}, F. and {Fraile}, E. and {Franke}, F. and {Gai}, M. and {Garabato}, D. and {Garcia-Gutierrez}, A. and {Garc{\'\i}a-Torres}, M. and {Garofalo}, A. and {Gavras}, P. and {Gerlach}, E. and {Geyer}, R. and {Giacobbe}, P. and {Gilmore}, G. and {Girona}, S. and {Giuffrida}, G. and {Gomel}, R. and {Gomez}, A. and {Gonzalez-Santamaria}, I. and {Gonz{\'a}lez-Vidal}, J.~J. and {Granvik}, M. and {Guti{\'e}rrez-S{\'a}nchez}, R. and {Guy}, L.~P. and {Hauser}, M. and {Haywood}, M. and {Helmi}, A. and {Hidalgo}, S.~L. and {Hilger}, T. and {H{\l}adczuk}, N. and {Hobbs}, D. and {Holland}, G. and {Huckle}, H.~E. and {Jasniewicz}, G. and {Jonker}, P.~G. and {Juaristi Campillo}, J. and {Julbe}, F. and {Karbevska}, L. and {Kervella}, P. and {Khanna}, S. and {Kochoska}, A. and {Kontizas}, M. and {Kordopatis}, G. and {Korn}, A.~J. and {Kostrzewa-Rutkowska}, Z. and {Kruszy{\'n}ska}, K. and {Lambert}, S. and {Lanza}, A.~F. and {Lasne}, Y. and {Le Campion}, J. -F. and {Le Fustec}, Y. and {Lebreton}, Y. and {Lebzelter}, T. and {Leccia}, S. and {Leclerc}, N. and {Lecoeur-Taibi}, I. and {Liao}, S. and {Licata}, E. and {Lindstr{\o}m}, E.~P. and {Lister}, T.~A. and {Livanou}, E. and {Lobel}, A. and {Madrero Pardo}, P. and {Managau}, S. and {Mann}, R.~G. and {Marchant}, J.~M. and {Marconi}, M. and {Marcos Santos}, M.~M.~S. and {Marinoni}, S. and {Marocco}, F. and {Marshall}, D.~J. and {Martin Polo}, L. and {Mart{\'\i}n-Fleitas}, J.~M. and {Masip}, A. and {Massari}, D. and {Mastrobuono-Battisti}, A. and {Mazeh}, T. and {McMillan}, P.~J. and {Messina}, S. and {Michalik}, D. and {Millar}, N.~R. and {Mints}, A. and {Molina}, D. and {Molinaro}, R. and {Moln{\'a}r}, L. and {Montegriffo}, P. and {Mor}, R. and {Morbidelli}, R. and {Morel}, T. and {Morris}, D. and {Mulone}, A.~F. and {Munoz}, D. and {Muraveva}, T. and {Murphy}, C.~P. and {Musella}, I. and {Noval}, L. and {Ord{\'e}novic}, C. and {Orr{\`u}}, G. and {Osinde}, J. and {Pagani}, C. and {Pagano}, I. and {Palaversa}, L. and {Palicio}, P.~A. and {Panahi}, A. and {Pawlak}, M. and {Pe{\~n}alosa Esteller}, X. and {Penttil{\"a}}, A. and {Piersimoni}, A.~M. and {Pineau}, F. -X. and {Plachy}, E. and {Plum}, G. and {Poggio}, E. and {Poretti}, E. and {Poujoulet}, E. and {Pr{\v{s}}a}, A. and {Pulone}, L. and {Racero}, E. and {Ragaini}, S. and {Rainer}, M. and {Raiteri}, C.~M. and {Rambaux}, N. and {Ramos}, P. and {Ramos-Lerate}, M. and {Re Fiorentin}, P. and {Regibo}, S. and {Reyl{\'e}}, C. and {Ripepi}, V. and {Riva}, A. and {Rixon}, G. and {Robichon}, N. and {Robin}, C. and {Roelens}, M. and {Rohrbasser}, L. and {Romero-G{\'o}mez}, M. and {Rowell}, N. and {Royer}, F. and {Rybicki}, K.~A. and {Sadowski}, G. and {Sagrist{\`a} Sell{\'e}s}, A. and {Sahlmann}, J. and {Salgado}, J. and {Salguero}, E. and {Samaras}, N. and {Sanchez Gimenez}, V. and {Sanna}, N. and {Santove{\~n}a}, R. and {Sarasso}, M. and {Schultheis}, M. and {Sciacca}, E. and {Segol}, M. and {Segovia}, J.~C. and {S{\'e}gransan}, D. and {Semeux}, D. and {Shahaf}, S. and {Siddiqui}, H.~I. and {Siebert}, A. and {Siltala}, L. and {Slezak}, E. and {Smart}, R.~L. and {Solano}, E. and {Solitro}, F. and {Souami}, D. and {Souchay}, J. and {Spagna}, A. and {Spoto}, F. and {Steele}, I.~A. and {Steidelm{\"u}ller}, H. and {Stephenson}, C.~A. and {S{\"u}veges}, M. and {Szabados}, L. and {Szegedi-Elek}, E. and {Taris}, F. and {Tauran}, G. and {Taylor}, M.~B. and {Teixeira}, R. and {Thuillot}, W. and {Tonello}, N. and {Torra}, F. and {Torra}, J. and {Turon}, C. and {Unger}, N. and {Vaillant}, M. and {van Dillen}, E. and {Vanel}, O. and {Vecchiato}, A. and {Viala}, Y. and {Vicente}, D. and {Voutsinas}, S. and {Weiler}, M. and {Wevers}, T. and {Wyrzykowski}, {\L}. and {Yoldas}, A. and {Yvard}, P. and {Zhao}, H. and {Zorec}, J. and {Zucker}, S. and {Zurbach}, C. and {Zwitter}, T.},
        title = "{Gaia Early Data Release 3. Summary of the contents and survey properties}",
      journal = {A\&A},
         year = 2021,
        month = may,
       volume = {649},
          eid = {A1},
        pages = {A1},
          doi = {10.1051/0004-6361/202039657},
archivePrefix = {arXiv},
       eprint = {2012.01533},
 primaryClass = {astro-ph.GA},
       adsurl = {https://ui.adsabs.harvard.edu/abs/2021A&A...649A...1G}
}

@dataset{Deep_GLIMPSE_ARCHIVE,
       author = {{Glimpse Team}},
        title = "{Deep GLIMPSE Archive}",
 howpublished = {NASA IPAC DataSet, IRSA201},
         year = 2020,
        month = jan,
          doi = {10.26131/IRSA201},
       adsurl = {https://ui.adsabs.harvard.edu/abs/2020ipac.data.I201G}
}

@misc{spherex_qr2_2025,
  author    = {{SPHEREx Team}},
  title     = {{SPHEREx Quick Release Spectral Images - QR2}},
  year      = {2025},
  publisher = {NASA/IPAC Infrared Science Archive},
  doi       = {10.26131/IRSA652},
  url       = {https://doi.org/10.26131/IRSA652}
}

@MISC{Deep_GLIMPSE_AAS,
       author = {{Whitney}, Barbara and {Benjamin}, Robert and {Churchwell}, Ed and {Meade}, Marilyn and {Babler}, Brian and {Allen}, Lori and {Anderson}, Loren and {Balser}, Dana and {Bania}, Thomas and {Blitz}, Leo and {Boyer}, Martha and {Brunt}, Chris and {Chakrabarti}, Sukanya and {Chambers}, Ed and {Clemens}, Dan and {Cohen}, Martin and {Cotera}, Angela and {Cyganowski}, Claudia and {Davis}, Chris and {Elmegreen}, Bruce and {Frinchaboy}, Peter and {Froebrich}, Dirk and {Hora}, Joseph and {Indebetouw}, Remy and {Ioannidis}, Georgios and {Jarrett}, Thomas and {Kerton}, Charles and {Kolbulnicky}, Henry and {Kraemer}, Kathleen and {Kumar}, Nanda and {Liu}, Sheng-Yuan and {Lucas}, Philip and {Majewski}, Steve and {Mauerhan}, Jon and {Marengo}, Massimo and {Megeath}, Tom and {Minniti}, Dante and {Mottram}, Joseph and {Povich}, Matthew and {Robitaille}, Thomas and {Rood}, Robert and {Sewilo}, Marta and {Smith}, Howard and {Smith}, Michael and {Stanke}, Thomas and {Stauffer}, John and {Van Dyk}, Schuyler and {van Loon}, Jacco and {Volk}, Kevin and {Watson}, Christer and {Wolf-Chase}, Grace and {Zasowski}, Gail},
        title = "{Deep GLIMPSE: Exploring the Far Side of the Galaxy}",
 howpublished = {Spitzer Proposal ID \#80074},
         year = 2011,
        month = may,
        pages = {80074},
       adsurl = {https://ui.adsabs.harvard.edu/abs/2011sptz.prop80074W}
}

@ARTICLE{NIRCam_performance,
       author = {{Rieke}, Marcia J. and {Kelly}, Douglas M. and {Misselt}, Karl and {Stansberry}, John and {Boyer}, Martha and {Beatty}, Thomas and {Egami}, Eiichi and {Florian}, Michael and {Greene}, Thomas P. and {Hainline}, Kevin and {Leisenring}, Jarron and {Roellig}, Thomas and {Schlawin}, Everett and {Sun}, Fengwu and {Tinnin}, Lee and {Williams}, Christina C. and {Willmer}, Christopher N.~A. and {Wilson}, Debra and {Clark}, Charles R. and {Rohrbach}, Scott and {Brooks}, Brian and {Canipe}, Alicia and {Correnti}, Matteo and {DiFelice}, Audrey and {Gennaro}, Mario and {Girard}, Julien H. and {Hartig}, George and {Hilbert}, Bryan and {Koekemoer}, Anton M. and {Nikolov}, Nikolay K. and {Pirzkal}, Norbert and {Rest}, Armin and {Robberto}, Massimo and {Sunnquist}, Ben and {Telfer}, Randal and {Wu}, Chi Rai and {Ferry}, Malcolm and {Lewis}, Dan and {Baum}, Stefi and {Beichman}, Charles and {Doyon}, Ren{\'e} and {Dressler}, Alan and {Eisenstein}, Daniel J. and {Ferrarese}, Laura and {Hodapp}, Klaus and {Horner}, Scott and {Jaffe}, Daniel T. and {Johnstone}, Doug and {Krist}, John and {Martin}, Peter and {McCarthy}, Donald W. and {Meyer}, Michael and {Rieke}, George H. and {Trauger}, John and {Young}, Erick T.},
        title = "{Performance of NIRCam on JWST in Flight}",
      journal = {\pasp},
         year = 2023,
        month = feb,
       volume = {135},
       number = {1044},
          eid = {028001},
        pages = {028001},
          doi = {10.1088/1538-3873/acac53},
archivePrefix = {arXiv},
       eprint = {2212.12069},
 primaryClass = {astro-ph.IM},
       adsurl = {https://ui.adsabs.harvard.edu/abs/2023PASP..135b8001R}
}

@ARTICLE{tr_coadds,
       author = {{Meisner}, A.~M. and {Lang}, D. and {Schlegel}, D.~J.},
        title = "{Time-resolved WISE/NEOWISE Coadds}",
      journal = {\aj},
         year = 2018,
        month = aug,
       volume = {156},
       number = {2},
          eid = {69},
        pages = {69},
          doi = {10.3847/1538-3881/aacbcd},
archivePrefix = {arXiv},
       eprint = {1710.02526},
 primaryClass = {astro-ph.IM},
       adsurl = {https://ui.adsabs.harvard.edu/abs/2018AJ....156...69M}
}

@ARTICLE{GRIPS,
       author = {{Paladini}, Roberta and {Zucker}, Catherine and {Benjamin}, Robert and {Nataf}, David and {Minniti}, Dante and {Zasowski}, Gail and {Peek}, Joshua and {Carey}, Sean and {Allen}, Lori and {Alonso-Garcia}, Javier and {Alves}, Joao and {Anders}, Friederich and {Athanassoula}, Evangelie and {Beers}, Timothy C. and {Bird}, Jonathan and {Bland-Hwathorn}, Joss and {Brown}, Anthony and {Buder}, Sven and {Casagrande}, Luca and {Casey}, Andrew and {Cassisi}, Santi and {Catelan}, Marcio and {Chary}, Ranga-Ram and {Chene}, Andre-Nicolas and {Ciardi}, David and {Comeron}, Fernando and {Cohen}, Roger and {Dame}, Thomas and {Drimmel}, Ronald and {Fernandez Trincado}, Jose and {Finkbeiner}, Douglas and {Geisler}, Douglas and {Gennaro}, Mario and {Goodman}, Alyssa and {Green}, Gregory and {Hajdu}, Gergely and {Henderson}, Calen and {Hora}, Joseph and {Ivanov}, Valentin D. and {Kirkpatrick}, Davy and {Kobayashi}, Chiaki and {Kuhn}, Michael and {Kunder}, Andres and {Lu}, Jessica and {Lucas}, Philip W. and {Majaess}, Daniel and {Megeath}, S. Thomas and {Meisner}, Aaron and {Molinari}, Sergio and {Mroz}, Przemek and {Ness}, Meliss and {Neumayer}, Nadine and {Nogueras-Lara}, Francisco and {Noriega-Crespo}, Alberto and {Poleski}, Radek and {Rix}, Hans-Walter and {Rebull}, Luisa and {Reggiani}, Henrique and {Rejkuba}, Marina and {Saito}, Roberto K. and {Schoenrich}, Ralph and {Saydjari}, Andrew and {Schisano}, Eugenio and {Schlafly}, Edward and {Schlaufman}, Keving and {Smith}, Leigh and {Speagle}, Joshua and {Wisz}, Dan and {Wyse}, Rosemary and {Zakamska}, Nadia},
        title = "{Roman Early-Definition Astrophysics Survey Opportunity: Galactic Roman Infrared Plane Survey (GRIPS)}",
      journal = {arXiv e-prints},
         year = 2023,
        month = jul,
          eid = {arXiv:2307.07642},
        pages = {arXiv:2307.07642},
          doi = {10.48550/arXiv.2307.07642},
archivePrefix = {arXiv},
       eprint = {2307.07642},
 primaryClass = {astro-ph.GA},
       adsurl = {https://ui.adsabs.harvard.edu/abs/2023arXiv230707642P}
}

@data{Line_LOWZ,
author = {Line, Michael},
publisher = {Harvard Dataverse},
title = {{LOWZ model atmosphere spectra}},
UNF = {UNF:6:ISBUKb9guFFBM6pEDYXoew==},
year = {2021},
version = {V1},
doi = {10.7910/DVN/SJRXUO},
url = {https://doi.org/10.7910/DVN/SJRXUO}
}
\bibliographystyle{aasjournalv7}

\end{document}